\documentclass[prd,showpacs,superscriptaddress]{revtex4-1}
\usepackage{amsmath}
\usepackage{graphicx,psfrag}
\usepackage{array}

\begin{document}
\title{Static weak dipole moments of the $\tau$ lepton via renormalizable scalar leptoquark interactions}

\author{A. Bola\~nos}
\affiliation{Facultad de
Ciencias F\'\i sico Matem\'aticas, Benem\'erita Universidad
Aut\'onoma de Puebla, Apartado Postal 1152, Puebla, Pue., M\'
exico}
\author{A. Moyotl}
\affiliation{Facultad de
Ciencias F\'\i sico Matem\'aticas, Benem\'erita Universidad
Aut\'onoma de Puebla, Apartado Postal 1152, Puebla, Pue., M\'
exico}
\author{G. Tavares-Velasco}
\email[Email: ]{gtv@fcfm.buap.mx}
\affiliation{Facultad de
Ciencias F\'\i sico Matem\'aticas, Benem\'erita Universidad
Aut\'onoma de Puebla, Apartado Postal 1152, Puebla, Pue., M\'
exico}

\begin{abstract}
The weak dipole moments of elementary fermions are calculated at the one-loop level in the framework
of a renormalizable scalar leptoquark model that forbids baryon number violating processes  and so
is free from the strong constraints from experimental data. In this model there are two scalar
leptoquarks  accommodated in an $SU_L(2)\times U_Y(1)$ doublet: one of such leptoquarks is
non-chiral and has electric charge of $5/3e$, whereas the other one is chiral and has electric
charge $2/3 e$. In particular, a non-chiral leptoquark contributes to the weak properties of an up
fermion via  a chirality flipping-term proportional to the mass of the virtual fermion and  can also
induce a non-zero weak electric dipole moment provided that the leptoquark couplings are complex.
The numerical analysis is focused on the weak properties of the $\tau$ lepton since they offer good
prospects for their experimental study. The constraints on leptoquark couplings are briefly
discussed for a non-chiral leptoquark with non-diagonal couplings to the second and third  fermion
generations, a third-generation non-chiral leptoquark, and a third-generation chiral leptoquark. It
is found that although the chirality-flipping term can enhance the weak properties of the $\tau$
lepton via the top quark contribution, such an enhancement would be offset by the strong constraints
on the leptoquark couplings. So, the contribution of scalar leptoquarks to the weak magnetic dipole
moment  of the $\tau$ lepton are smaller than the standard model (SM) contributions but can be of
similar size than those arising in some SM extensions. A
non-chiral leptoquark can also give contributions to the weak electric dipole moment larger than the SM
one but well below the experimental limit. We also  discuss the case of  the
off-shell  weak dipole moments and  for completeness   analyze  the behavior of the
$\tau$ electromagnetic properties.

\end{abstract}
\pacs{}
\maketitle

\section{Introduction}
\label{Introduction}
The study of the static electromagnetic  properties of charged leptons has long played a central
role in  experimental particle physics. The   magnetic dipole moment (MDM) and the
electric dipole moment (EDM), which can only arise for spinning particles, have drawn
as much attention as that devoted to other particle properties. Although the electron MDM, $a_e$, has been an
instrumental probe of quantum electrodynamics,   any new physics
contribution to $a_e$ is too small to be at the reach of detection, so the experimental measurements
are commonly employed to determine the value of the fine structure constant rather than to look for
evidences of new physics. A different scenario arises in the case of  the muon  MDM, $a_\mu$, which
receives sizeable contributions from all the sectors of the standard model (SM). Even more,
$a_\mu$ can be determined with a very high precision both experimentally and theoretically and thus it has become  a powerful benchmark to test
the SM with very high accuracy and to search for  effects of physics beyond the SM.  The most recent experimental determination of $a_\mu$, which has reached a precision of 0.7 parts per million
\cite{Bennett:2006fi}, leads to a discrepancy with the SM prediction at the
level of 3.6 standard deviations:
\begin{equation}
\label{discamu}
\Delta{a_\mu}=a_\mu^{\rm Exp.}-a_\mu^{\rm SM}= 287(80) \times 10^{-11},
\end{equation}
where the experimental and theoretical errors have been added in quadrature. Although such a
discrepancy may be a signal of new physics,  a more accurate
calculation of the hadronic light-by-light contribution is yet to be obtained.

On the other hand,  our knowledge of the $\tau$ lepton electromagnetic properties is still
unsatisfactory, which stems from the fact that the $\tau$ lifetime is very short to prevent  its
interaction with an electromagnetic field from direct measurements. The most stringent current bound on
$a_\tau$ with 95 \% C.L., $-0{.}052< a_\tau< 0{.}013$, was obtained by looking for deviations from the SM in  the cross section of
the process $e^+e^-\to e^+e^-\tau^+\tau^-$ using  the data collected by the DELPHI collaboration at
the CERN large electron positron (LEP2) collider during the years 1997-2000 \cite{Abdallah:2003xd}, while the theoretical SM prediction is $a_\tau^{\text{SM}}=1177{.}2
1(5)\times10^{-6}$ \cite{Eidelman:2007sb}.  It turns out that a precise measurement of the tau
MDM is required as it could confirm or rule out  the possibility that the $\Delta a_\mu$
discrepancy  is a signal  of new physics: the natural scaling of heavy particle effects on a lepton
MDM implies that $\Delta a_\tau/\Delta a_\mu\sim m_\tau^2/m_\mu^2$, so if the current $a_\mu$
discrepancy is interpreted as a new physics effect,  we would expect that $\Delta a_\tau \simeq
10^{-6}$. Although the SM prediction disagrees with this value,  some of its extensions, such as the
SeeSaw model \cite{Biggio:2008in}, the minimal supersymmetric standard model with a mirror fourth
generation\cite{Ibrahim:2008gg}, and unparticle physics \cite{Moyotl:2012zz}, predict that $a_\tau$
lies in the interval of $10^{-6}$ to $10^{-10}$.

As for the EDM, it represents a useful tool for the study of the discrete symmetry $CP$ and
provides a potential probe to unravel its origin. However, the only
EDMs that can be directly measured are those  of the neutron, the proton, the deuteron and the muon,
whereas the EDM of other particles can only be indirectly determined. Although an experimental signal of an EDM is yet to be detected, the
best current upper limits on the electron and muon EDMs come from the study of the  thallium EDM
\cite{Regan:2002ta} and the E821 experiment at Brookhaven \cite{Bennett:2008dy}, respectively:
 \begin{eqnarray}
d_e &<&(6.9\pm7.4)\times10^{-28} \,\,e\text{cm},\\
d_\mu &<&(-0.1\pm0.9)\times10^{-19}\,\,e \text{cm},
\end{eqnarray}
whereas the experimental detection of  the $\tau$ EDM  poses the same difficulties
as its MDM.  Nevertheless, the
$\tau$ EDM was searched for in the $e^+e^-\to\tau^+\tau^-$ reaction by the Belle collaboration
\cite{Inami:2002ah} at the KEK collider. The achieved  sensitivity, in units of $10^{-16}\,\,e$cm,
was
\begin{eqnarray}
-0{.}22<&  \text{Re}(d_\tau)& < 0{.}45,\\
-0{.}25 < &\text{Im}(d_\tau) &< 0{.}08.
\end{eqnarray}
In the SM, the  EDM of a lepton is predicted to be negligibly small as it  arises at the three-loop level of perturbation theory, which   can be a blow for  its experimental detection. However, several SM extensions predict sizeable contributions  that can be at the experimental reach. Although the $\tau$ short lifetime represents a  challenge, this  lepton emerges as a natural candidate to search for new physics effects such as a large EDM because of its mass and  wide spectrum of decay channels.

The study of $\tau$ physics plays a significant role in $B$ factories. For instance, the
ill-fated Super$B$ accelerator with its 75 ab$^{-1}$ was expected to measure
the $\tau$ EDM  with a resolution of
$|\text{Re}(d_\tau)|=7.2\times10^{-20}$ $e$cm \cite{GonzalezSprinberg:2007qj}, whereas the expected
resolution of the real and imaginary parts of $a_\tau$ was estimated to be of the order of
$(0.75-1.7)\times10^{-6}$ \cite{Bernabeu:2007rr,*O'Leary:2010af}. With a
lower planned luminosity, the upgraded  Belle II facility at the KEK B-factory will offer unique
perspectives for the study of $\tau$  physics in both high precision measurements of the SM
parameters and new physics searches. The electromagnetic dipole moments of the $\tau$ lepton  may be
measured via the radiative leptonic decays $\tau^- \to \ell ^-\nu_\tau \bar{\nu}_\ell \gamma$
($\ell=e,\mu$) \cite{Laursen:1983sm}. However, this method is only sensitive to large values of
$a_\tau$, so a more detailed analysis will determine the feasibility of this
proposal.

Contrary to the attention drawn to  the static electromagnetic properties of fermions, a lot of work is still necessary to have a better understanding of their weak properties, namely the $CP$-conserving  weak magnetic dipole moment (WMDM) and the $CP$-violating weak electric dipole moment (WEDM), which are
the coefficients of dimension-five operators in the effective Lagrangian of the $\bar{f}fZ$ interaction and
can   be extracted from  the following terms of the respective vertex function
\begin{eqnarray}
\Gamma^\mu_Z(q^2)&=&F_2\left(q^2\right)i\sigma^{\mu\nu}
q_\nu+F_3\left(q^2\right)\sigma^{\mu\nu} \gamma_5 q_\nu, \label{lagew}
\end{eqnarray}
where $q=p_2-p_1$ is the $Z$ gauge boson transferred four-momentum. The WMDM, $a_f^W$, and
the WEDM, $d_f^{W}$, are defined at the $Z$-pole: $a_f^W=-2m_f
F_2\left(m_Z^2\right)$ and $d_f^W=-eF_3\left(m_Z^2\right)$. Since $a_f^W$ and $d_f^W$  are the coefficients of chirality-flipping terms, they are expected to
give contributions proportional to some positive power of the mass of the involved fermion. This
allows one to construct  observable quantities that can be experimentally proved, but which
are particularly suited for  heavy fermions, among from which the $\tau$ lepton,  the $b$ quark, and
the $t$ quark are the most promising candidates. In particular, a large value of the WEDM of charged
fermions  would lead to a considerable deviation of the total $Z$ width  from its SM value
\cite{Bernreuther:1988jr}, which can provide an indirect upper limit on the corresponding WEDM.
Along these lines, the study of the $Z\to\tau^+\tau^-$ decay at  center-of-mass energies near the
$Z$ resonance represents a promising tool to search for signals of the weak dipole moments of  the $\tau$
lepton. This process could allow one to measure  $a_\tau^W$ and $d_\tau^W$  through the
transverse and normal polarizations of the $\tau$ leptons \cite{Bernabeu:1993er}. By following this
approach, the ALEPH collaboration obtained the current best limit on the $\tau$ WMDM and WEDM, with
95$\%$ C.L. \cite{Heister:2002ik}:

\begin{eqnarray}
\text{Re}(a_\tau^{W}) &<& 1{.}1\times10^{-3},\\
\text{Im}(a_\tau^{W})&<&  2{.}7 \times 10^{-3},\\
\text{Re}(d_\tau^{W}) &<& 0{.}5 \times 10^{-17}\,\,e\text{cm},\\
\text{Im}(d_\tau^{W})&<&  1{.}1 \times 10^{-17}\,\,e\text {cm},
\end{eqnarray}
which were extracted from the data collected at the CERN from 1990 to 1995, corresponding to an
integrated luminosity of 155 pb${}^{-1}$. These bounds are far above the SM predictions
$a_\tau^W=-(2.10+0.61i)\times10^{-6}$ \cite{Bernabeu:1994wh} and $d_\tau^W<8\times10^{-34}
\,\,e$cm \cite{Bernreuther:1988jr}. However, the CERN large hadron collider  (LHC) could open a door for the experimental study of
these properties. Along these lines, a study of the $pp\to \tau^+\tau^-$
and $pp\to Zh\to\tau^+\tau^-h$ cross sections including  anomalous $Z$ couplings was presented in
Ref. \cite{Hayreter:2013vna}. It was found that an analysis at the LHC would allow
experimentalists to
measure the deviations from the SM and extract constraints on the $\tau$ electromagnetic and weak
dipole moments.

The weak properties of a fermion have been studied in several SM extensions, such as models allowing
tree-level flavor changing neutral currents \cite{Queijeiro:1994kb}, the two-Higgs-doublet model  (THDM)
\cite{Bernabeu:1995gs,*GomezDumm:1999tz}, the minimal supersymmetric standard model (MSSM)
\cite{deCarlos:1997br,*Hollik:1997vb}, the minimal supersymmetric version of the SM with complex
parameters  \cite{Hollik:1997ph}, and in the context of unparticle physics \cite{Moyotl:2012zz}.
Prompted by a recent work \cite{Arnold:2013cva} on the study of the simplest renormalizable scalar leptoquark models with no proton decay (see also Ref. \cite{Dorsner:2009cu}), we will consider  one of such models for our study. This model is interesting as there is a non-chiral leptoquark that could give rise to large contributions to the weak properties of a charged lepton due to a chirality-flipping term. Furthermore, very recently it was shown that such a scalar leptoquark, with mass below 1 TeV, can provide an explanation for the observed branching ratios of the $B\to D^{*}\tau \bar{\nu}$ decays \cite{Dorsner:2013tla}.
The rest of this article is organized as
follows. A brief review on the scalar leptoquark model we are interested in along with the details
of the calculation of the weak properties of a fermion is presented in Sec. \ref{Calculation}.
Section \ref{Analysis} is devoted to the numerical analysis of our results, with particular
emphasis to the $\tau$ lepton  weak properties, including a short discussion on the case of the off-shell dipole moments.  For completeness we will also discuss the $\tau$ electromagnetic properties. The conclusions and outlook
are presented in Sec. \ref{Conclusions}.

\section{Weak dipole moments of a fermion in scalar leptoquark models}
\label{Calculation}

The vanishing of gauge anomalies in the SM due to the interplay of charged fermions hints to a profound link between lepton and quarks in a more fundamental theory, such as the one conjectured long ago by Pati and Salam \cite{Pati:1974yy}, which gives rise to new leptoquark particles carrying both lepton  and baryon numbers. Leptoquark particles can be of scalar or vector type and are also predicted in grand unified theories (GUTs) \cite{Georgi:1974sy,*Senjanovic:1982ex,*Frampton:1989fu}, composite models \cite{Schrempp:1984nj},  technicolor models \cite{Farhi:1980xs,*Hill:2002ap}, superstring-inspired ${\rm E}_6$ models \cite{Witten:1985xc,*Hewett:1988xc}, etc. For a more comprehensive listing of this class of models along with low-energy constraints, the reader may want to refer to  \cite{Davidson:1993qk,*Hewett:1997ce}.  Of special interest are leptoquark models in which baryon and lepton numbers are individually conserved, thereby forbidding any tree-level contribution to proton decay induced by leptoquark couplings to diquarks. As a result, in such models the leptoquark mass can be as light as the electroweak scale, which contrast with some GUT-inspired leptoquark models in which it must lie around the Planck scale in order to avoid a rapid proton decay.

Because of the complexity inherent to leptoquark models, it has been customary to analyze their potential effects in a model-independent fashion via the effective lagrangian approach: the most general dimension-four  ${\rm SU}_c(3)\times {\rm SU}_L(2)\times {\rm U}_Y(1)$ invariant lagrangian
parameterizing both scalar and vector leptoquark couplings
satisfying both baryon and lepton number conservation was first presented in
\cite{Buchmuller:1986zs}. Quite recently, the authors of Ref. \cite{Arnold:2013cva} brought the
attention to the only two minimal renormalizable models where scalar leptoquarks are introduced via
a single representation of ${\rm SU}_c(3)\times {\rm SU}_L(2)\times {\rm U}_Y(1)$ and in which there is no
proton decay induced via tree-level  leptoquark exchange. This fact was also stressed previously in Ref. \cite{Dorsner:2009cu}.  We will focus on one of this models,
dubbed Model I in Ref. \cite{Arnold:2013cva}, and calculate the corresponding contribution
  to the WMDM and the WEDM of a fermion. Although the  WEDM of heavy fermions has already been studied in the context of leptoquark models \cite{Bernreuther:1996dr,Poulose:1997kt}, to our knowledge there is no previous analysis on the behavior of the contributions of leptoquarks to the WMDM. In the model I of Ref. \cite{Arnold:2013cva} there is a non-chiral leptoquark
(it has both left- and right-handed couplings) that  gives rise to a chirality-flipping contribution
to the  electromagnetic and weak properties of a fermion.
Such a term is proportional to the internal quark mass and it is worth examining if there is an enhancement in the  contribution from a heavy internal fermion.

In the model we are interested in, there is a scalar leptoquark doublet $R_2^T=(R_{1/2},R_{-1/2})$ with  quantum numbers $(3; 2; 7/6)$ under the ${\rm SU}_c(3)\times {\rm SU}_L(2)\times {\rm U}_Y(1)$ gauge group. For our calculation we only need to consider the following zero-fermion-number effective interaction \cite{Buchmuller:1986zs,Mizukoshi:1994zy}:

\begin{equation}
{\cal L}_{F=0}= h^{ij}_{2L}R_2^T \bar{u}^i_Ri\tau_2 \ell^j_L+h^{ij}_{2R}\bar{q}^i_Le^j_R R_2+{\rm H.c.},
\end{equation}
with ${\ell_L^i}^T=(\nu_L^i,e_L^i)$ and ${q_L^i}^T=(u_L^i,d_L^i)$.

From the above Lagrangian we obtain the interaction of a lepton-quark pair with two  scalar leptoquarks $S_1\equiv R_{1/2}$ and $S_2\equiv R_{-1/2}$, with electric charge of $5/3e$ and $2/3e$. We write the interaction Lagrangian as
\begin{equation}
\label{LagF=0}
{\cal L}_{F=0}=\bar{e}^i\left(\lambda_{L}^{ij}P_L+ \lambda_{R}^{ij}P_R\right)u^j S_1^*+\bar{e}^i \eta_R^{ij} P_R d^jS_2^*+{\rm H.c.},
\end{equation}
where $P_{L,R}$ are the chiral projection operators and  $i,j$ are generation indices.  The flavor eigenstates were rotated to the mass eigenstates and so the  $\lambda_{L}^{ij}$, $\lambda_{R}^{ij}$, and $\eta^{ij}_R$ couplings already encompass this information.  Notice that while $S_1$ has both left- and right-handed couplings to a charged lepton and an up quark, $S_2$  has only right-handed couplings.

For our purpose we will also need the leptoquark couplings with the photon
and the $Z$ boson, which are extracted from the leptoquark kinetic Lagrangian and can be written in
the form
\begin{equation}
{\cal L}= ieQ_{S_k}
{S_k}\overleftrightarrow{\partial_\mu} S_k^*  A^\mu- \frac{ig}{c_W}g_{Z{S_k}{S_k}} {S_k}
\overleftrightarrow{\partial_\mu} S_k^* Z^\mu+{\rm H.c.},
\end{equation}
where $g_{ZS_1S_1}=1/2-s_W^2 Q_S$ and $g_{ZS_2S_2}=-1/2-s_W^2 Q_S$. Here $Q_i$ is the electrical charge in units of $e$. For completeness, we present the SM interactions of the photon and
the $Z$ gauge boson with a fermion pair:
\begin{equation}
{\cal L}=i eQ_i\bar{f}_i\gamma_\mu f_i
A^\mu -\frac{ig}{c_W}\bar{f}_i\gamma_\mu\left(g^{i}_L P_L+g^{i}_R P_R\right) f_i Z^\mu.
\end{equation}
In particular, we will need below $g^u_L=\frac{1}{2}-\frac{2s_W^2}{3}$ and $g^u_R=\frac{s_W^2}{3}$ for the contribution of an up quark to the $\tau$ weak properties. The corresponding Feynman rules can be extracted straightforwardly from the above Lagrangians.

\begin{figure}[!ht]
\begin{center}
\includegraphics[width=4in]{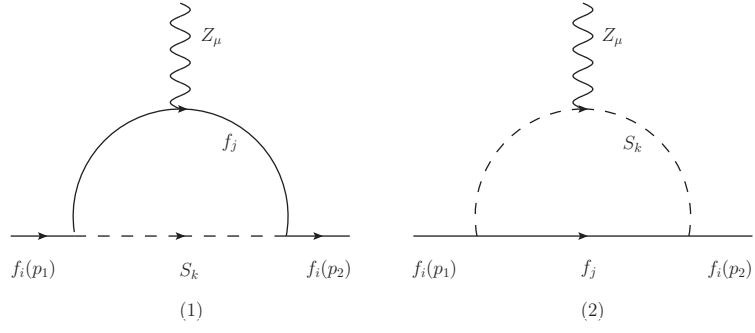}
\end{center}
\caption{Feynman diagrams inducing the fermion weak dipole moments via a scalar leptoquark $S_k$.  When the fermion $f_i$ is a lepton (quark), the internal fermion is a quark (lepton). Bubble diagrams do not contribute to the dipole moments.
\label{FeynmanDiagrams}}
\end{figure}
\subsection{Weak dipole moments of a fermion}

At the one-loop level, the weak properties of the fermion $f_i$  are induced by a scalar leptoquark and a fermion $f_j$ via the Feynman diagrams of Fig. \ref{FeynmanDiagrams}.  The method of Feynman parameters yields the following results

\begin{eqnarray}
\label{aWi}
a_{i}^W&=&-\frac{3\,\sqrt{x_{i}}}{32\pi^2s_Wc_W}\Bigg[\sqrt{x_{i}}\left(
\left|\lambda^{ij}_L\right|^2 F_{L}({x_{i}},{x_{j}},x_Z)+ \left|\lambda^{ij}_R\right|^2
F_{R}({x_{i}},{x_{j}},x_Z)
\right)\nonumber\\&+&
2\sqrt{x_{j}}{\rm Re}\left(\lambda^{ij}_L {\lambda^{ij}_R}^*\right) G({x_{i}},{x_{j}},x_Z)\Bigg],
\end{eqnarray}
where we have defined  $x_A=m_A^2/m_{S_k}^2$ and
\begin{eqnarray}
F_{L,R}(z_1,z_2,z_3)&=&g_{L,R}^{j}F_1(z_1,z_2,z_3)+g_{{S_k}{S_k}Z}F_2(z_1,z_2,z_3),\\
G(z_1,z_2,z_3)&=&\frac{1}{2}\left(g_{L}^{j}+g_{R}^{j}\right)G_1(z_1,z_2,z_3)+g_{{S_k}{S_k}Z}G_2(z_1,z_2,z_3).
\end{eqnarray}
The $F_a$ and $G_a$
functions, which stand for the contributions of each one of the Feynman diagram of Fig. \ref{FeynmanDiagrams}, can be written as
\begin{equation}
\label{Faweak}
F_{a}(z_1,z_2,z_3)=8\int_0^1 \frac{(1-x) x}{\sqrt{  z_3
\chi_a(x,z_1,z_2,z_3)}}\arctan
\left(\frac{\sqrt{z_3}\xi_a(x)}{\sqrt{\chi_a(x,z_1,z_2,z_3)}}\right)dx,
\end{equation}
and
\begin{equation}
\label{Gaweak}
G_a(z_1,z_2,z_3)=8 \int_0^1 \frac{(1-x)}{\sqrt{  z_3
\chi_a(x,z_1,z_2,z_3)}}\arctan\left(\frac{\sqrt{z_3}\xi_a(x)}{\sqrt{\chi_a(x,z_1,z_2,z_3)}}\right)dx,
\end{equation}
with  the auxiliary functions $\chi$, $\eta$ and $\xi_i$ defined by $\chi(x,z_1,z_2,z_3)=4\eta(x,z_1,z_2)-z_3\xi^2_a(x) $,
$\eta(x,z_1,z_2)=(1-x) (z_2-xz_1)+x$, $\xi_1(x)=1-x$, and $\xi_2(x)=x$.

As for the weak electric dipole moment of fermion $f_i$, it can be written in the form
\begin{eqnarray}
\label{dWi}
d_{i}^W=\frac{3\, e \sqrt{x_{i}x_{j}} }{32m_{i}s_Wc_W \pi^2}{\rm
Im}\left(\lambda^{ij}_L {{\lambda^{ij}_R}}^*\right)  G({x_{i}},{x_{j}},x_Z).
\end{eqnarray}
As mentioned above,  there is a chirality-flipping term proportional to the internal
fermion mass when the leptoquark is non-chiral. Below, we will concentrate on the potential effects of such a leptoquark on the weak properties of the $\tau$ lepton as there can be an important enhancement from the $t$ quark contribution.

\subsection{Electromagnetic dipole moments of a fermion}

For the completeness of our analysis we will also need the contributions from scalar leptoquarks to the static
electromagnetic properties of a fermion, which follow easily from our calculation by taking the
limit  $m_Z \to 0$ and replacing the $Z$ couplings by the photon ones. It can be helpful to test the validity of our results. We thus obtain the scalar
leptoquark contribution to the magnetic dipole
moment $a_{i}$ and the electric dipole moment $d_{i}$ of  fermion $f_i$:

\begin{eqnarray}
\label{afi}
a_{i}=-\frac{3\,\sqrt{x_{i}}}{32\pi^2} \left[\sqrt{x_{i}}\left(\left|\lambda^{ij}_L\right|^2+
\left|\lambda^{ij}_R\right|^2\right) F^\gamma({x_{i}},{x_{j}})+2\,\sqrt{x_{j}}\,{\rm Re}\left(\lambda^{ij}_L
{\lambda^{ij}_R}^*\right) G^\gamma({x_{i}},{x_{j}}) \right],
\end{eqnarray}
and
\begin{eqnarray}
\label{dfi}
d_{i}=\frac{3\, e \sqrt{x_{i}x_{j}} }{32m_{i}\pi^2}{\rm Im}\left(\lambda^{ij}_L
{\lambda^{ij}_R}^*\right)  G^\gamma({x_{i}},{x_{j}}),
\end{eqnarray}
where
\begin{eqnarray}
F^\gamma(z_1,z_2,z_3)&=&Q_j F^\gamma_1(z_1,z_2,z_3)+Q_{S_k} F^\gamma_2(z_1,z_2,z_3),\\
G^\gamma(z_1,z_2,z_3)&=&Q_j G^\gamma_1(z_1,z_2,z_3)+Q_{S_k} G^\gamma_2(z_1,z_2,z_3),
\end{eqnarray}
with the $F_a^\gamma$ and $G_a^\gamma$ functions given by
\begin{equation}
\label{Fagfeyn}
F^\gamma_a(z_1,z_2)=2\int_0^1\frac{(1-x) x \xi_a(x)}{(1-x)(z_2- x z_1)+x}dx,
\end{equation}
\begin{equation}
\label{Gagfeyn}
G^\gamma_a(z_1,z_2)=2\int_0^1\frac{(1-x) \xi_a(x)}{(1-x)(z_2- x z_1)+x}dx.
\end{equation}
The equations can be integrated explicitly in the limit of a very heavy leptoquark, in which case we obtain

\begin{eqnarray}
\label{Fgamma1}
F^\gamma_1(0,z)= \frac{1}{3(1-z)^4}\left(2+3z-6z^2+ z^3+6 z \log
(z)\right),\\
\label{Fgamma2}
F^\gamma_2(0,z)= \frac{1}{3(1-z)^4}\left(1-6z+3z^2+ 2z^3-6 z^2 \log (z)\right),
\end{eqnarray}
\begin{eqnarray}
\label{Ggamma1}
G^\gamma_1(0,z)=-\frac{1}{(1-z)^3}\left(3-4z+z^2+2 \log (z)\right),\\
\label{Ggamma2}
G^\gamma_2(0,z)=\frac{1}{(1-z)^3}\left(1-z^2+2 z
\log (z)\right).
\end{eqnarray}
These  results are in agreement with previous results for the magnetic \cite{Djouadi:1989md,Cheung:2001ip}
and the electric \cite{Bernreuther:1990jx} dipole moments induced by scalar leptoquarks. In
addition,  we present in Appendix \ref{PassarinoResults} an  alternative calculation of the weak
and electromagnetic properties of a fermion in terms of Passarino-Veltman scalar functions, which
can be used to make a cross-check of our results.

\section{Numerical analysis}
\label{Analysis}
\subsection{Leptoquark constraints}

In the following analysis we will concentrate on the
$\tau$ lepton electromagnetic and weak properties as  they offer
good prospects for their experimental study. We will consider a charge $5/3e$ non-chiral scalar leptoquark as  it is expected to give the dominant contribution to the electromagnetic and weak properties of the $\tau$ lepton in the model we are considering.
The phenomenology of such a leptoquark has been studied considerably in the past
\cite{Shanker:1982nd,Davidson:1993qk,Mizukoshi:1994zy} and very recently  \cite{Arnold:2013cva, Dorsner:2013tla}
with constraints from the $Z\to b\bar{b}$ decay, the muon MDM,
lepton flavor violating decays  and the
$\tau$ EDM. There are strong constraints from low energy physics \cite{Shanker:1982nd,Davidson:1993qk,Mizukoshi:1994zy} on
leptoquarks that couple to the first-generation fermions, so we
will assume  a leptoquark that only has non-negligible couplings to fermions of the second and third generations. As for the leptoquark mass, the
most stringent constraint on the mass of a third-generation chiral scalar leptoquark, $m_S> 526$
GeV, was obtained from the analysis of the data from the LHC \cite{Chatrchyan:2012st}. It was  assumed
that such a leptoquark decays mainly into a bottom quark and a $\tau$ lepton,
such as occurs with the  $S_2$ leptoquark.  Since it is required that $S_1$ and $S_2$ are mass
degenerate or have a small mass splitting to avoid large contributions to the oblique parameters
\cite{Keith:1997fv}, we will assume a  leptoquark with a mass larger than  500 GeV.

Below we will focus on three illustrative scenarios and discuss briefly the constraints on the leptoquark
coupling constants to present a realistic analysis.  We will then analyze the behavior of
the $\tau$ electromagnetic and weak properties as a function of
the leptoquark mass and also discuss the case of the off-shell dipole moments.

\subsubsection{Scenario I: a non-chiral leptoquark with non-diagonal couplings to the second and
third fermion generations}

We first analyze the scenario in which there exists a non-chiral leptoquark that can have complex
non-diagonal couplings to lepton-quark pairs of the second and third generations. In this scenario,
the $CP$-even electromagnetic and weak properties of the $\tau$ lepton receive the contributions of a non-chiral leptoquark accompanied by the $c$ or the $t$ quarks and can be  enhanced by the
chirality-flipping term.  In addition, there can be non-zero CP-violating
properties. On the negative side, such a leptoquark can give rise to large contributions to the muon
 MDM and the LFV decay
$\tau\to \mu\gamma$, which in turn can impose strong constraints on the leptoquark couplings. We note that a similar scenario is posed by a scalar singlet leptoquark, such as the one whose behavior was analyzed in
\cite{Benbrik:2008si}, which can also be non-chiral but it is known to give dangerous contributions
to the proton decay via its diquark couplings.

\begin{figure}[!ht]
\begin{center}
\includegraphics[width=4in]{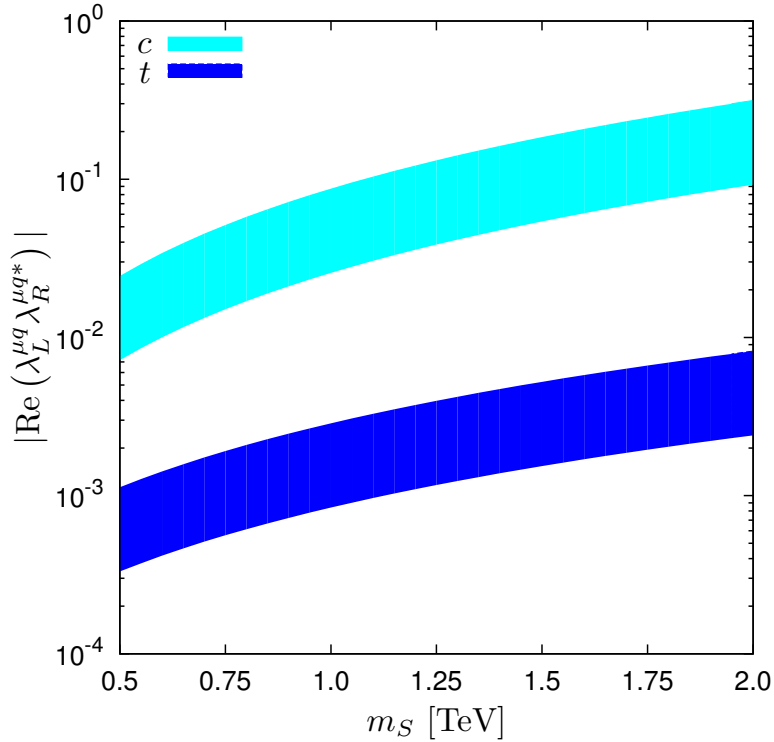}
\end{center}
\caption{Allowed areas with 95 \% C.L. on the $m_S$ vs  $|{\rm Re}\left(\lambda_{L}^{\mu
q} {\lambda_{R}^{\mu q}}^*\right)|$  plane consistent with the current experimental limit
on the muon MDM.  The light-shaded (dark-shaded) area corresponds to the allowed region for the
contribution of
the $c$ ($t$) quark. It is assumed that either the $c$ or the $t$ quark contribution is responsible
for the $a_\mu$ discrepancy. Notice that the left- and right-handed leptoquark
couplings must have opposite signs in order to give a positive contribution to the muon MDM.
\label{boundamu}}
\end{figure}

If one assumes that the discrepancy on the muon MDM (\ref{discamu}) is due entirely to our scalar leptoquark, the
allowed region for its respective contribution, with 95\% C.L, is $130.2\times 10^{-11} \le
\Delta a^{\rm NP}_\mu\le 443.8 \times 10^{-11}$.  We assume that either the $c$
or the $t$ quark contribution is responsible for the $a_\mu$ discrepancy and obtain the allowed
regions on the $m_S$ vs $|{\rm Re}\left(\lambda_{L}^{\mu q} {\lambda_{R}^{\mu q}}^*\right)|$   plane
with 95 \% C.L., which we  show in Fig. \ref{boundamu}.
Although the product ${\rm Re}\left(\lambda_{L}^{\mu t} {\lambda_{R}^{\mu t}}^*\right)$ is tightly
constrained, a less stringent constraint would be obtained if the  $t$ quark contribution was
canceled out by the $c$ quark contribution or another new physics contribution.

\begin{figure}[!ht]
\begin{center}
\includegraphics[width=4in]{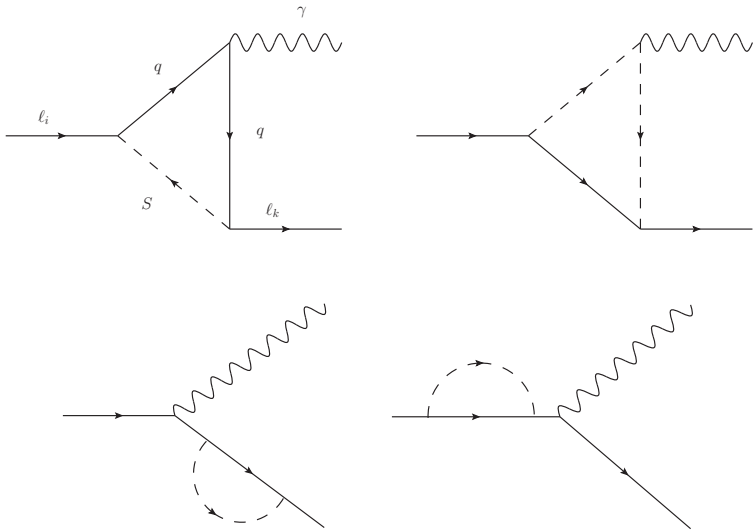}
\end{center}
\caption{Feynman diagrams for the radiative decay $\ell_i \to \ell_k \gamma$ induced by at the
one-loop level by
quarks and scalar leptoquarks.
\label{litolkg}}
\end{figure}

If the leptoquark has non-diagonal couplings to the second and third generations, the LFV $\ell_i
\to
\ell_k \gamma$ decay can proceed via the Feynman diagrams of Fig. \ref{litolkg}. The contribution of
 leptoquark $S_k$ and quark $q$  to the $\ell_i
\to
\ell_k \gamma$  decay
amplitude can be written in the form
\begin{equation}
\label{amplilkgamma}
{\cal M}(\ell_i \to \ell_k \gamma)=-\frac{ie}{m_{S_k}}\bar{\ell}_k(p_k)
\left[A_L({x_{i}},x_{q},{x_{k}})P_L+A_R({x_{i}},x_{q},{x_{k}})
P_R\right]\sigma_{\mu\nu}\ell_i(p_i)q^{\nu}\epsilon^\mu(q),
\end{equation}
where the $A_L$ and $A_R$ coefficients are given by
\begin{eqnarray}
\label{ALlilkg}
A_{L}(z_1,z_2,z_3)&=&\frac{3}{16\pi^2}\left[\lambda_{L}^{kq}
{\lambda^{iq}_L}^*\sqrt{z_1}I(z_1,z_2,z_3) +\lambda_{R}^{kq} {\lambda^{qi}_R}^*
\sqrt{z_3}I(z_3,z_2,z_1)+\lambda_{L}^{kq} {\lambda^{iq}_R}^* \sqrt{z_2} J(z_1,z_2,z_3) \right],
\end{eqnarray}
and $A_R=A_L\left(L\leftrightarrow R\right)$. This decay was already studied in
\cite{Povarov:2011zz}, but  we have made our own evaluation  for completeness and we present  the
$I$ and $J$ functions in Appendix
\ref{decaylitoljgamma} in terms of Feynman parameter integrals
and  Passarino-Veltman scalar functions. The respective decay width is  given by
\begin{equation}
\Gamma(\ell_i \to \ell_k
\gamma)=\frac{m_i x_{i}\alpha}{4}\left[1-\frac{x_k}{x_i}
\right]^3\left(|A_L(x_i,x_q,x_k)|^2+|A_R(x_i,x_q,x_k)|^2\right).
\end{equation}

As far as the experimental constraints on LFV decays are concerned,
quite recently the upper bound on the decay rate for the $\mu \to e\gamma$  decay was improved up to
$5.7\times 10^{-13}$ by the MEG collaboration \cite{Adam:2013mnn}, but the bounds on the
LFV $\tau$ decays are much weaker: BR$(\tau\to e \gamma) < 3.3 \times 10^{-8}$ and ${\rm BR}(\tau\to
\mu \gamma) < 4.4 \times 10^{-8}$ \cite{Aubert:2009ag}. Since both the $\mu \bar{q}S$ and
$\tau \bar{q} S$ vertices, with
$q=c,t$, enter into the amplitude of the $\tau\to \mu\gamma$ decay, it can only be
useful to constrain the product $\lambda^{\mu q}\lambda^{\tau q}$, so a bound on $\lambda^{\tau q}$
will be largely dependent on
$\lambda^{\tau q}$.

\begin{figure}[!ht]
\begin{center}
\includegraphics[width=4in]{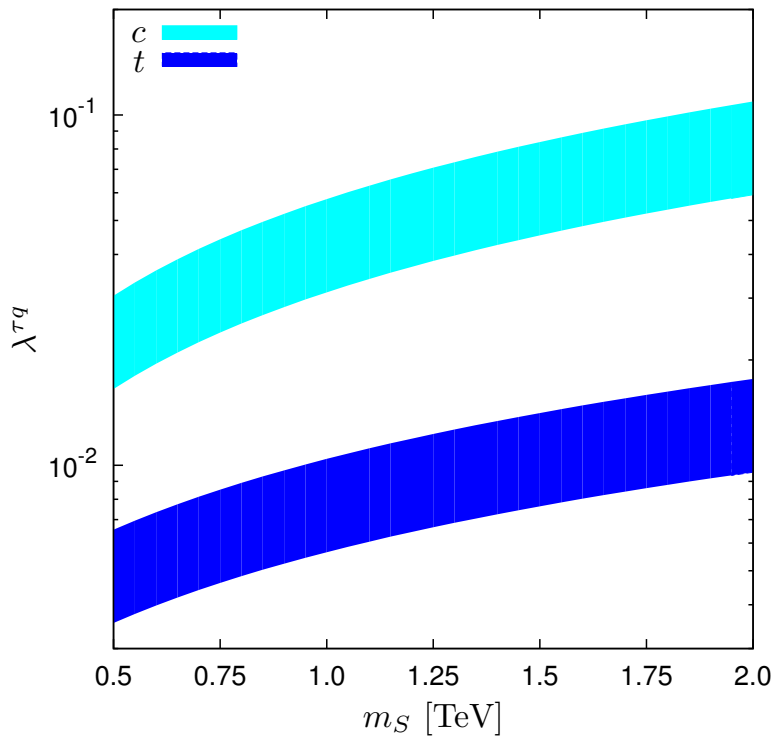}
\end{center}
\caption{Allowed area with 95 \% C.L. on the $m_S$ vs $\lambda^{\tau q}$   plane
consistent with the experimental limits on the muon MDM and the $\tau\to \mu\gamma$ decay. We
considered  $|\lambda_{L}^{\ell q}|=|\lambda_{R}^{\ell q}|\equiv  \lambda^{\ell q}$
($\ell=\mu,\tau$ and $q=c,t$), with $\lambda^{\mu q}$ lying
inside the allowed area shown of Fig. \ref{boundamu}. The light-shaded (dark-shaded) region is the
allowed area for the $c$  quark ($t$ quark) contribution.
\label{boundlfv}}
\end{figure}

For simplicity we take  $|\lambda_{L}^{\ell q}|=|\lambda_{R}^{\ell q}|\equiv \lambda^{\ell q}$
($\ell=\mu,\tau$) and
consider that  either the $c$ quark or the $t$ quark contribution is the only responsible for the
$a_\mu$ discrepancy, i.e. we assume
that the $\lambda^{\mu q}$, coupling takes on values inside the allowed area
shown in Fig.
\ref{boundamu}. We then  obtain the plot of  Fig.
\ref{boundlfv}, where we show the allowed region on the $m_S$ vs $\lambda^{\tau q}$ plane
consistent with both the experimental constraints on the muon MDM and the LFV decay $\tau\to
\mu\gamma$. We observe that, in order to explain the $a_\mu$ discrepancy and be consistent with the
$\tau\to \mu\gamma$ decay, the $\lambda^{\tau c}$ coupling must reach values of the order of about
$10^{-1}$, whereas $\lambda^{\tau t}$ must reach values one order of magnitude smaller.
A word of caution is on order here, the
allowed areas of Fig. \ref{boundlfv} would alter  if the $q$ quark
contribution was not assumed to be the only responsible for the $a_\mu$ discrepancy, which in
turn could be explained by other
contributions of the same model.

\subsubsection{Scenario II: a third-generation  non-chiral leptoquark}
This scenario is similar to the first scenario  except that there are  no leptoquark-mediated LFV
processes and therefore the constraints on the leptoquark couplings are less stringent than in
scenario I. The electromagnetic and weak properties of the $\tau$ lepton, which would only receive the contribution of the non-chiral leptoquark and the $t$ quark, can still be  enhanced by the chirality-flipping term and there can be  non-zero $CP$-violating properties provided
that the leptoquark couplings are complex.  Therefore, this scenario can provide the largest
values of the electromagnetic and weak properties of the $\tau$ lepton. Constraints on the
couplings of such a  leptoquark were obtained in Ref.  \cite{Mizukoshi:1994zy} by performing a
global fit to the  LEP data on $Z$ physics. It was found that leptoquark couplings of the order of about $10^{-1}$
are allowed provided that the leptoquark mass is of the order of  $600$ GeV.

\subsubsection{Scenario III: a third-generation chiral leptoquark}
Although a chiral scalar leptoquark with couplings to fermions of the second-generation would yield a
negative contribution  to the muon MDM, which is disfavored by the experimental data, such  a
contribution  would vanish for a third-generation leptoquark. In this case the only contributions to the  electromagnetic and weak properties of the $\tau$ lepton  arise from the chiral leptoquark accompanied by the $t$ quark. Apart that this scenario does not induce $CP$ violating properties, it appears to be
unfavorable for large values of the  electromagnetic and weak properties  as they would be
naturally suppressed due to the absence of the chirality-flipping term.  Constraints on this class of leptoquarks were obtained  in Ref.
\cite{Bhattacharyya:1994ig} from the experimental measurement of the  partial decays $Z\to \ell^+\ell^-$. It was found
that a third-generation leptoquark with a mass larger than about 500 GeV and a coupling to the $t$
quark and the $\tau$ lepton of  electroweak strength $g$ are compatible.

\subsection{Behavior of the electromagnetic and weak properties of the $\tau$ lepton}

In general, the $f\bar{f}V$ vertex is gauge invariant and gauge independent only when the gauge
boson is on its mass shell, therefore the pinch technique was used in
\cite{Papavassiliou:1993qe} to construct gauge-independent electromagnetic and weak
dipole form factors.  It was argued \cite{Denner:1994nn}, however, that  off-shell form factors are
not uniquely defined and  so they do not represent observable quantities. In the case of the
leptoquark contribution to the weak and electromagnetic dipole moments, there are no internal gauge
bosons circulating in the loops and so there is no dependence on the gauge-fixing parameter.  Our
results for the weak dipole moments can thus be easily generalized for  arbitrary squared momentum
$q^2\equiv s$ of the gauge boson by replacing $x_Z\to s/m_{S_k}^2$ in Eqs.  (\ref{aWi}) and (\ref{dWi}). The
electromagnetic dipole form factors follow easily after exchanging the $Z$ couplings by the photon ones.
The resulting quantities can be useful to assess the sensitivity to the effects of leptoquark
particles on the dipole form factors, as suggested in the analysis presented in \cite{Bernabeu:1995gs} within the framework of the THDM. Below we will analyze the $\tau$ electromagnetic properties
for $\sqrt{s}=0$ and $\sqrt{s}=10.5$ GeV. On the other
hand, the $\tau$ weak properties will be analyzed for $\sqrt{s}=m_Z$ and $\sqrt{s}=500$ GeV. The values of $\sqrt{s}$ used for the off-shell gauge bosons are the center-of-mass energies of a $B$ factory and the future next linear collider, respectively. Furthermore, in our study below we will consider the
interval $10^{-2}-10^{-1}$  for $\lambda^{\tau q}$ and $500$ GeV -- $2000$ GeV for $m_S$,
which are in accordance with the bounds from experimental data discussed above. Here $\lambda^{\tau
q}$ represents the leptoquark coupling constants.

\begin{figure}[!ht]
\begin{center}
\includegraphics[width=6.in,height=7.5in]{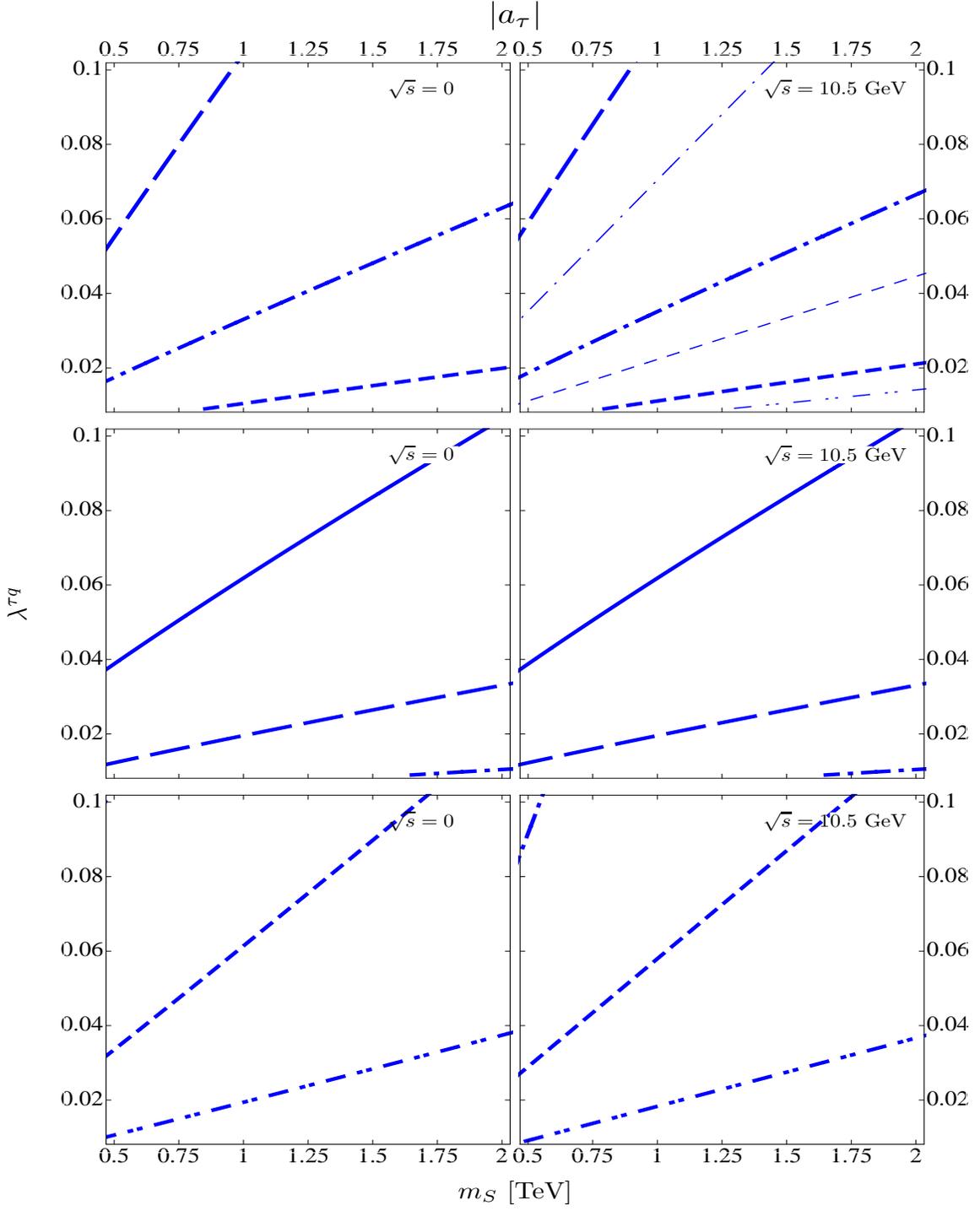}
\end{center}
\caption{
Contours  of  $a_\tau$ in the $m_S$ vs $\lambda^{\tau q}$ plane for a nonchiral leptoquark
accompanied
by the $c$ quark (upper plots), a nonchiral leptoquark accompanied by the $t$ quark (middle plots)
and a third-generation chiral scalar
leptoquark (lower plots). The thick (thin) lines represent the
contours of the absolute value of the real (imaginary) part of $a_\tau$. The  values of the
contours are $10^{-7}$ (full lines), $10^{-8}$
(long-dashed lines),  $10^{-9}$ (dash-dotted lines), $10^{-10}$ (short-dashed line),
and  $10^{-11}$ (dash-dot-dotted lines).
For the nonchiral leptoquark   $\lambda^{\tau q}=\lambda^{\tau q}_L=\lambda^{\tau q}_R$ whereas for
the chiral
leptoquark we use either $\lambda^{\tau q}=\lambda^{\tau q}_L$ and $\lambda^{\tau q}_R=0$ or
$\lambda^{\tau q}=\lambda^{\tau q}_R$ and
$\lambda^{\tau q}_L=0$. \label{taumdm}}
\end{figure}

We first assume that the leptoquark  couplings are real and calculate the following leptoquark
contributions to the $\tau$ MDM: that of a non-chiral leptoquark with non-diagonal couplings to a
lepton-quark pair of the second and third families and that of a third-generation chiral leptoquark.
In the former case there are contributions from the $c$ and the $t$ quarks, but in the latter
there is only a contribution from the $t$ quark. We show in Fig. \ref{taumdm} the contours of the $\tau$
MDM in the $m_S$ vs $\lambda^{\tau q}$ plane for $\sqrt{s}=0$ and $\sqrt{s}=10.5$ GeV, with
$\lambda^{\tau q}=\lambda^{\tau q}_L\ne 0$ and $\lambda^{\tau q}_R=0$ for the chiral leptoquark (i.e. a
left-handed leptoquark), and $\lambda^{\tau q}=\lambda^{\tau q}_L=\lambda^{\tau q}_R$ for the
non-chiral
leptoquark. Note that
the MDM is insensitive to the chirality of the leptoquark, so our results are valid for either a
left- or a right-handed leptoquark.  As mentioned
above,   there is an enhancement of the non-chiral leptoquark contribution due to the presence of
the chirality-flipping term, but it would be significant only in the case of  the $t$ quark, whose
contribution can reach  values slightly above the $10^{-8}$ level for $\lambda^{\tau t}\sim 10^{-2}$
and up to $10^{-7}$-$10^{-6}$ for $\lambda^{\tau t}\sim 10^{-1}$. However, in the case of
the $c$ quark, a small value of the coupling constant would offset the enhancement from  the
chirality-flipping term and this  contribution would be below the $10^{-9}$ level for $\lambda^{\tau
c}\simeq 10^{-2}$.   In the case  of a third-generation chiral leptoquark,  although  its  couplings
were of the order of $10^{-1}$, its contributions would be
lower than  the contribution of a non-chiral
leptoquark accompanied by the $t$ quark, provided that the couplings of the non-chiral leptoquark
are of the
order of $10^{-2}$. In general, the contribution of a chiral leptoquark is slightly
dependent on the quark mass, which is due to the absence of the
chirality-flipping
term. If the photon goes off-shell, with $\sqrt{s}=10.5$ GeV, there is a slight increase
in the real part of  $a_\tau$ and at the same time an imaginary part is developed in the case of the
$c$ quark
contribution since $\sqrt{s}>2m_c$. Such an imaginary part would be slightly smaller than the
corresponding real part.

\begin{figure}[!ht]
\begin{center}
\includegraphics[width=6.in,height=7.5in]{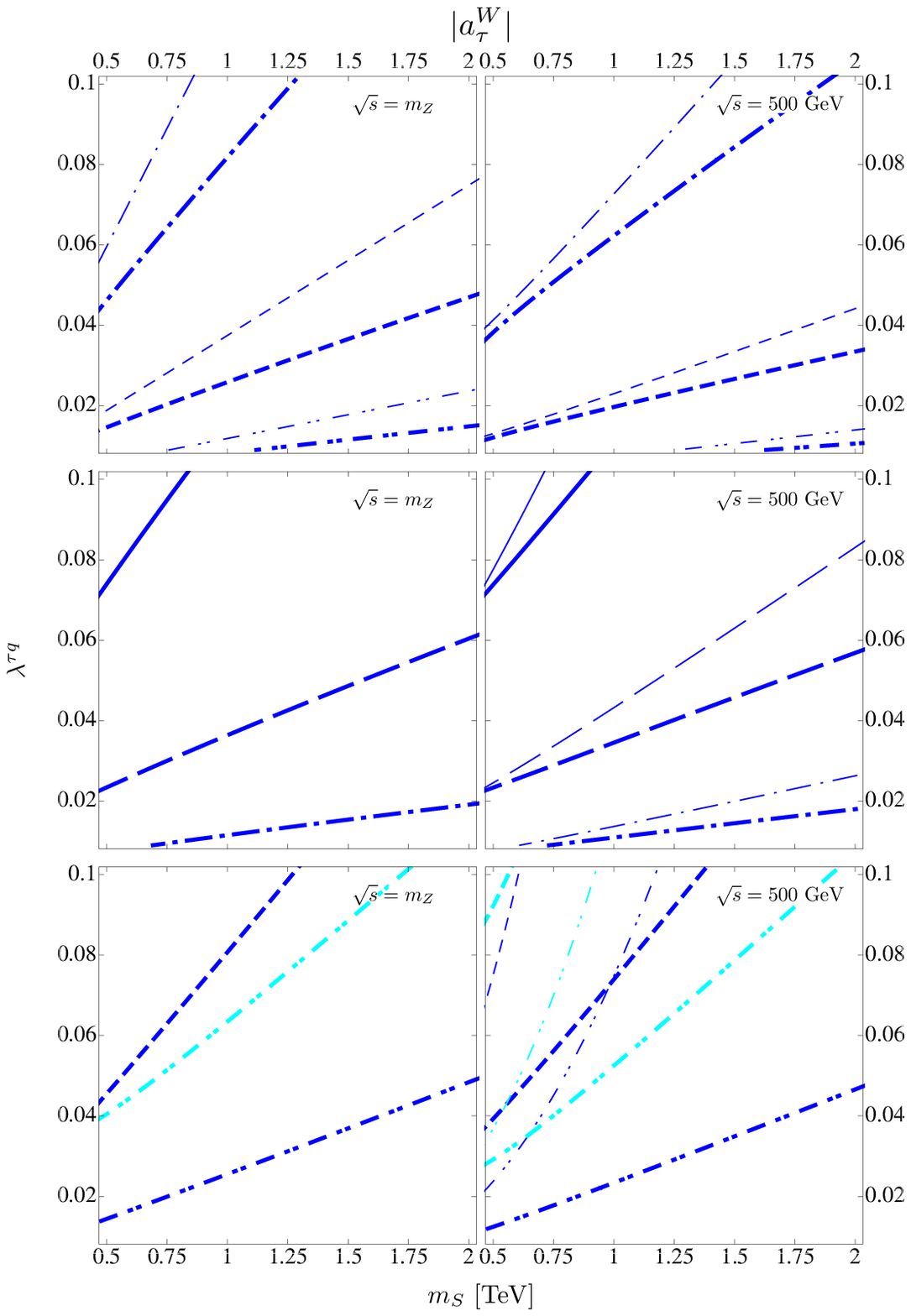}
\end{center}
\caption{Contours  of  $a^W_\tau$ in the $m_S$ vs $\lambda^{\tau q}$ plane for a nonchiral
leptoquark
accompanied
by the $c$ quark (upper plots), a nonchiral leptoquark accompanied by the $t$ quark (middle plots),
a left-handed third-generation chiral scalar
leptoquark (darker lines of the lower plots) and a right-handed third-generation chiral scalar
leptoquark (lighter lines of the lower plots). The thick (thin) lines represent the
contours of the absolute value of the real (imaginary) part of $a^W_\tau$ and  the values of the
contours are $10^{-7}$ (full lines), $10^{-8}$
(long-dashed lines),  $10^{-9}$ (dash-dotted lines), $10^{-10}$ (short-dashed line),
and  $10^{-11}$ (dash-dot-dotted lines).
For the nonchiral leptoquark  $\lambda^{\tau q}=\lambda^{\tau q}_L=\lambda^{\tau q}_R$ whereas for
the left-handed and right-handed
leptoquark $\lambda^{\tau q}=\lambda^{\tau q}_L$  and $\lambda^{\tau q}=\lambda^{\tau q}_R$,
respectively.  \label{tauwmdm}}
\end{figure}

We now show the contours of the $a^W_\tau$ in the $m_S$ vs $\lambda^{\tau q}$ plane   in Fig.
\ref{tauwmdm} for $\sqrt{s}=m_Z$ and $\sqrt{s}=500$ GeV. Again we show the contributions of a
non-chiral leptoquark accompanied by the $c$ or the $t$ quarks and, since the WMDM is sensitive to
the chirality of the leptoquark,  we now show the contributions of both  a left- and a right-handed
leptoquark.  The largest contribution to the static $a^W_\tau$ would arise from a
non-chiral leptoquark accompanied by the $t$ quark, which can reach the level of $10^{-9}$ for
$\lambda^{\tau t}=10^{-2}$ and $10^{-7}$ for $\lambda^{\tau t}=10^{-1}$, whereas the contribution of
a non-chiral leptoquark and the $c$ quark is about two orders of magnitude smaller. The latter
contribution now develops an imaginary part that is slightly smaller than the real part. As far as
the third-generation chiral leptoquark is concerned, the contribution of a left-handed leptoquark
can be as large as $10^{-10}$ for $m_S\simeq 500$ GeV and $\lambda^{\tau t}_L\simeq 10^{-1}$ but it
is much  smaller than the contribution of a non-chiral leptoquark and the $t$ quark for smaller
$\lambda^{\tau q}_L$ and larger $m_S$. On the other hand, the contribution of a right-handed
leptoquark is about one order of magnitude smaller than that of a left-handed one. In general
the enhancement due to the chirality-flipping term, which appears only in the non-chiral leptoquark
contribution, is less pronounced in the case of the WMDM than in the case of the MDM.
When the $Z$ gauge boson goes off-shell, the real part of the leptoquark
contributions to $a^W_\tau$ shows a rather similar behavior to that observed in the case of an on-shell $Z$ gauge boson, but
 all the contributions develop  an imaginary part, which is smaller than the
corresponding real part. Although there is an increase in the magnitude of the contributions when the
$Z$ gauge boson goes off-shell, with the largest increase observed in the contribution of a non-chiral
leptoquark accompanied by the $c$ quark, such an increase is moderate. To summarize, the
largest contribution to $a^W_\tau$ would arise from a third-generation non-chiral leptoquark even if
its couplings were one order of magnitude below than those of a third-generation chiral leptoquark.

\begin{figure}[!ht]
\begin{center}
\includegraphics[width=6in,height=5.25in]{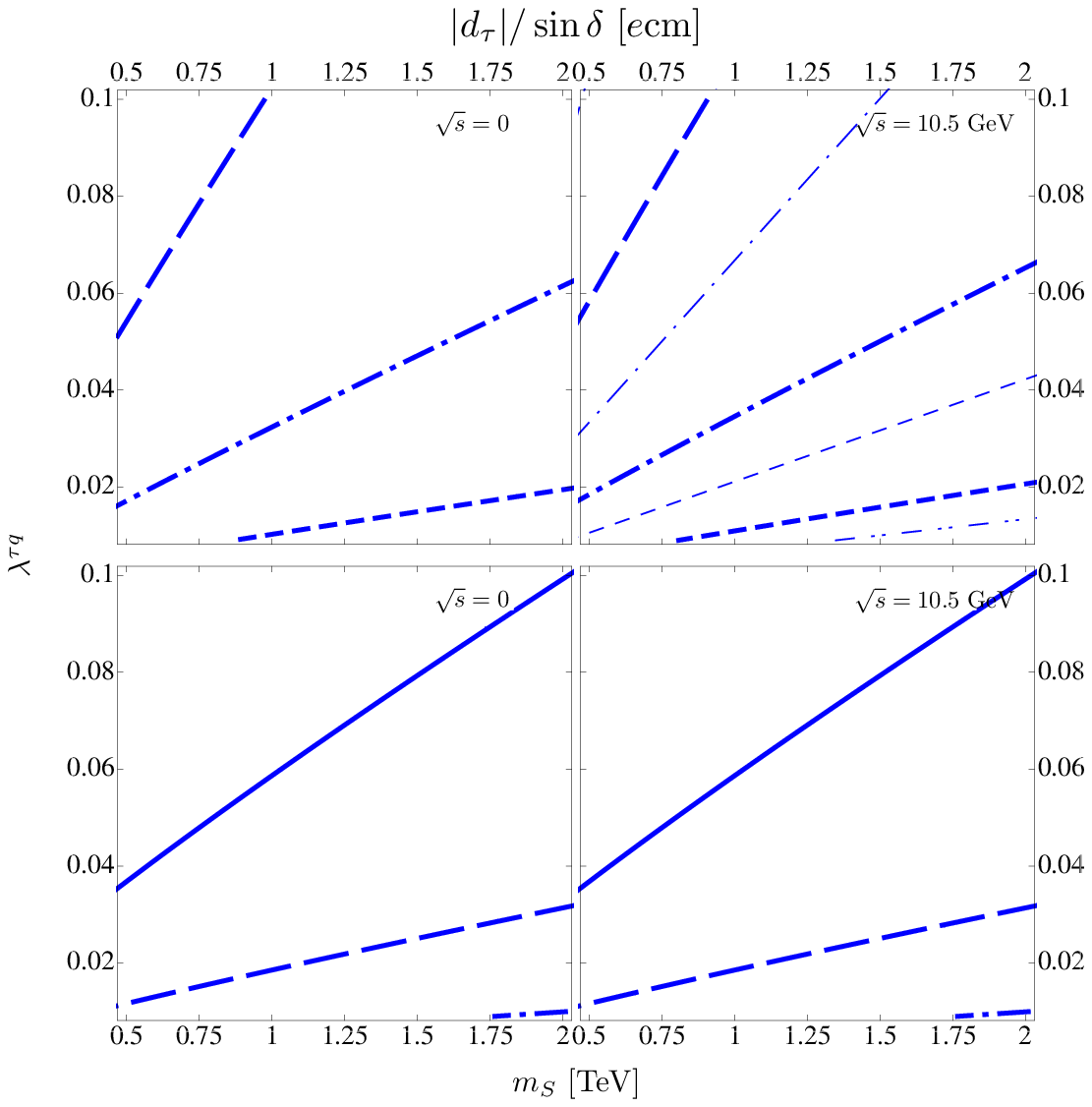}
\end{center}
\caption{Contours  of  $d_\tau/\sin\delta$ in the $m_S$ vs $\lambda^{\tau q}$ plane for a nonchiral
leptoquark
accompanied
by the $c$ quark (upper plots) and the $t$ quark (lower plots). The thick (thin) lines represent the
contours of the absolute value of the real (imaginary) part of $d_\tau$. The values of the
contours are $10^{-21}$ (full lines), $10^{-22}$
(long-dashed lines),  $10^{-23}$ (dash-dotted lines), $10^{-24}$ (short-dashed line),
and  $10^{-25}$ (dash-dot-dotted lines), in units of $e$cm. We set  $\lambda^{\tau q}=\lambda^{\tau
q}_L=\lambda^{\tau q}_R$.
\label{tauedm}}
\end{figure}

\begin{figure}[!ht]
\begin{center}
\includegraphics[width=6in,height=5.25in]{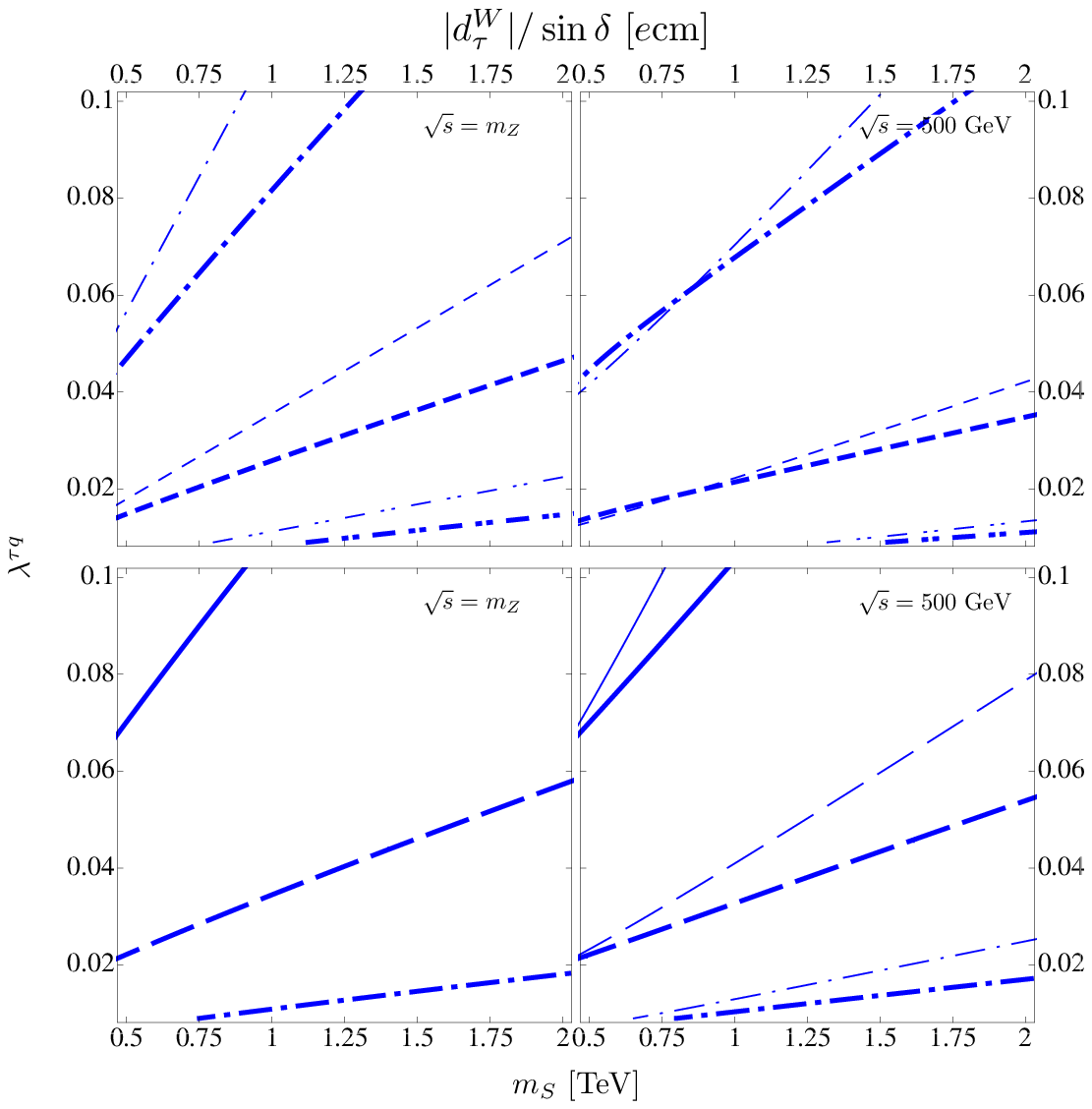}
\end{center}
\caption{The same as in Fig. \ref{tauedm} but for  $d^W_\tau/\sin\delta$.
 \label{tauwedm}}
\end{figure}

We now turn to analyze the $CP$-violating properties of the $\tau$ lepton, which can only arise when
the leptoquark is non-chiral and its couplings are complex. These
properties are proportional to  $\sin\delta_q$, with $\delta_q$ the relative phase between the
$\lambda^{\tau q}_L$ and $\lambda^{\tau q}_R$ coupling constants. We first show in
Fig.
\ref{tauedm} the contours of the contribution  to the $\tau$ EDM arising from a non-chiral scalar leptoquark accompanied by the
$c$ or the $t$ quark in the $m_S$ vs $\lambda^{\tau q}$ plane, for $\sqrt{s}=0$ and
$\sqrt{s}=10.5$ GeV. We observe that, irrespectively of the value of $\sqrt{s}$, the $t$ quark
contribution to
$d_\tau$ can be as large as $10^{-23}$-$10^{-20}$ $e$cm for $\lambda^{\tau t}\sim 10^{-1}$ and
$m_S\sim 500$ GeV, but
it is two
orders of magnitude below for $\lambda^{\tau t}\sim 10^{-2}$ and $m_S\sim 2000$ GeV. On the other
hand, the
$c$ quark contribution is much smaller and  can only reach the level of $10^{-22}$ $e$cm even if
$\lambda^{\tau c}\sim 10^{-1}$.  Apart from a slight increase in the real part of $d_\tau$, the only
noticeable difference between the contributions
of an
on-shell and an off-shell photon is the imaginary part that is developed by the $c$ quark
contribution, which is smaller than corresponding real part.
Finally, we analyze the leptoquark contribution to the WEDM of the $\tau$ lepton, which also
is non-vanishing only for a non-chiral leptoquark with complex couplings. In Fig. \ref{tauwedm} we
show the contours of the contribution of such a leptoquark to $d^W_\tau/\sin\delta_q$ in the $m_S$ vs
$\lambda^{\tau q}$
plane for $\sqrt{s}=m_Z$ and $\sqrt{s}=500$ GeV. As expected, the $t$ quark yields the
dominant contribution, with values ranging between $10^{-23}$ to $10^{-21}$ $e$cm, whereas
the $c$ quark contribution is much smaller. When $\sqrt{s}=500$ GeV, the behavior of the real part
of the contribution of the $t$ quark remains almost unchanged with respect to the case of an
on-shell $Z$ gauge boson, though an imaginary part is developed. A more pronounced change is
observed in the behavior of the contribution of the $c$ quark, which can reach larger values as
$\sqrt{s}$ increases, which is evident by the downward shift of the contour lines as larger values of $d^W_\tau$ can be reached for smaller values of $\lambda^{\tau q}$.

In summary,  the $\tau$ static electromagnetic and weak properties induced by a scalar leptoquark
can
reach the values  shown in Table \ref{estimates} in the scenarios discussed above, considering
values for the coupling constants consistent with the constraints from experimental data. For
comparison purpose, we also include the predictions of other extensions of the SM. The reader is referred to the original references for the particular values of the parameters used to obtain these estimates. Notice that there can be additional suppression in these values as we show the largest ones we can expect in every model. In the case of
scenario I, although the electromagnetic and weak properties can be enhanced by the contribution
of the $t$ quark, such an enhancement would likely be offset since the
leptoquark couplings are strongly constrained. On the other hand, although the constraints on the
leptoquark couplings are less stringent  in scenario III  than in
scenarios I and II,  the electromagnetic and weak properties of  the $\tau$ lepton
are naturally suppressed in such a scenario due to the absence of a chirality-flipping term. In conclusion, scenario II seems to
be the one
that can give rise to the largest values of the $\tau$ electromagnetic and weak properties as the
constraints on the leptoquark couplings are less stringent than in scenario I. Furthermore, in this
scenario there can be nonzero
$CP$-violating properties, which are absent in scenario III. The respective contributions would be,
however, smaller than in other SM models such as the MSSM. For an off-shell photon or $Z$ gauge
boson there is no appreciable difference in the order of magnitude of the real part of the
electromagnetic and weak dipole moments, though depending on the value of $s$ an
imaginary part can develop in the case of the
$c$ quark contribution to the electromagnetic properties and the $t$ contribution to the weak dipole
moments. Such an imaginary part is absent in the case of on-shell gauge bosons. It is worth mentioning that, for a very heavy
scalar leptoquark, the only difference between the  MDM and the WMDM of distinct
charged leptons would arise from  the actual value of the coupling constants since there would be no
appreciable difference
arising from the numerical values of the loop functions due to the small values of the  lepton
masses.

A comment is in order here regarding previous evaluations of the $CP$-violating electromagnetic and
weak properties of the $\tau$ lepton induced by scalar leptoquarks.
The authors of Ref. \cite{Bernreuther:1996dr} present  expressions for $d_\tau$ and $d^W_\tau$ at
arbitrary $s$ obtained via the Passarino-Veltman method. We have checked that there is
agreement between those results and the ones presented in Appendix \ref{PassarinoResults} after the
replacement $x_Z\to s/m_{S_k}^2$ is done. On the other hand, the authors of  Ref. \cite{Poulose:1997kt}
present integral formulas for the imaginary and real part of $d_\tau$ and $d^W_\tau$, which were
obtained via the Cutkosky rules. In these works  the $CP$-violating dipole moments are numerically evaluated for
leptoquark coupling constants  of the order of unity or larger and a
leptoquark mass  below $500$ GeV. Although we do not consider that region of the parameter space
since we present an up-to-date analysis using
parameter values that are still in accordance with current experimental data in order to obtain a
realistic estimate of  the electromagnetic and weak properties of the $\tau$ lepton,
we have verified that our results agree numerically with those presented in the aforementioned
works. Furthermore,
as stated before, to our knowledge, there is no previous analysis of the
leptoquark contribution to the WMDM of the $\tau$ lepton.

\setlength{\tabcolsep}{6pt}
\begin{table}[!ht]
\caption{Estimate for the contribution of a scalar leptoquark to the static electromagnetic and weak
properties of the $\tau$ lepton assuming a value of $10^{-2}$ ($10^{-1}$) for the non-chiral
(chiral) leptoquark couplings and  a leptoquark mass of $500$ GeV in the three scenarios
discussed in the text, where $\delta$ stands for the imaginary phase between the left- and
right-handed leptoquark couplings. When the photon or the $Z$ gauge
boson are off-shell there can be an increase
in the values shown in the table and also an absorptive part can develop. We also include the
predictions of the MSSM.
\cite{deCarlos:1997br,*Hollik:1997vb,Hollik:1997ph},
the THDM \cite{GomezDumm:1999tz,Bernabeu:1995gs,Bernabeu:1994wh}, and unparticle physics (UP)
\cite{Moyotl:2012zz}.\label{estimates}}
\begin{tabular}{ccccccc}
\hline\\
Scenario &$a_\tau$ &$d_\tau$\; [$e$cm] &${\rm
Re}\left(a^W_\tau\right)$ &${\rm
Im}\left(a^W_\tau\right)$ &${\rm Re}\left(d^W\right)_\tau$\;
[$e$cm] &${\rm
Im}\left(d^W_\tau\right)$\; [$e$cm]\\[0.2cm]
\hline
I,II&$10^{-8}$&$10^{-22}\times \sin\delta$&$10^{-9}$&$10^{-10}$&$10^{-22}\times
\sin\delta$&$10^{-24}\times \sin\delta$\\
%\hline
%II&&&&&&\\
\hline
III&$10^{-9}$&--&$10^{-10}$&--&--&--\\
\hline
MSSM &$10^{-6}$\cite{Ibrahim:2008gg}&$10^{-18}$ \cite{Ibrahim:2010va}&$10^{-6}$ \cite{Hollik:1997vb}&$10^{-7}$ \cite{Hollik:1997vb}&$10^{-21}$\cite{Hollik:1997ph}&--\\
\hline
THDM &$10^{-6}$&$10^{-24}$ \cite{GomezDumm:1999tz}&$10^{-10}$\cite{Bernabeu:1994wh}&--&$10^{-22}$\cite{GomezDumm:1999tz}&--\\
\hline
UP \cite{Moyotl:2012zz}&$10^{-6}$&$10^{-21}$&$10^{-9}$&$10^{-9}$&$10^{-24}$&$10^{-24}$\\
\hline
\end{tabular}
\end{table}

\section{Conclusions and final remarks}
\label{Conclusions}

We have calculated the static weak properties of a fermion induced by a scalar leptoquark motivated
by a recent work on the analysis of the simplest renormalizable scalar leptoquark models with no
proton decay \cite{Arnold:2013cva}. We consider one of such models, the one that predicts  a
non-chiral scalar leptoquark that can induce at the one-loop level the weak properties of the $\tau$
lepton, whose study is interesting as there are good prospects for their experimental study. For
completeness we also study the $\tau$ electromagnetic properties. We analyze three particular
scenarios and discuss the constraints on the leptoquark couplings to obtain a realistic estimate,
namely, we consider a non-chiral leptoquark with non-diagonal couplings to the second and third
generations, a third-generation non-chiral leptoquark, and a third-generation chiral leptoquark. In
the case of the non-chiral leptoquark there can be  a significant  enhancement due a
chirality-flipping term proportional to the top quark mass, but such term is absent in the case of a
chiral leptoquark and its contributions to the $\tau$ electromagnetic and weak properties are
naturally suppressed. However, the chirality-flipping  term  can also give rise to large
contributions to LFV processes and leptonic $Z$ decays, thereby imposing strong constraints on the
leptoquark couplings. Therefore, the enhancement given by the chirality-flipping term is partially
offset by the small value of the coupling constants. We find that the most promising scenario for
the largest contributions to the electromagnetic and weak properties of the $\tau$ lepton is that of
a third-generation non-chiral leptoquark, which can induce contributions to the MDM and WMDM of the
same order of magnitude than those predicted by  SM extensions such as the THDM, namely, ${\rm
Re}\left(a^W_\tau\right)\simeq 10^{-9}$ and ${\rm Im}\left(a^W_\tau\right)\simeq 10^{-10}$, though
these contributions are well below the SM ones. A non-chiral leptoquark can also contribute to the
$CP$-violating EDM and WEDM, namely ${\rm Re}\left(d^W_\tau\right)\simeq 10^{-22}$ $e$cm and ${\rm
Im}\left(d^W_\tau\right)\simeq 10^{-24}$ $e$cm, which are much larger than the SM values but still
far from the  experimental limits. In particular, the values of the leptoquark contribution to $d_\tau$ and $d^W_\tau$ are considerably smaller than the ones found in previous works since we consider values of the coupling constant and mass of the leptoquark consistent with current experimental data. We also analyzed
the scenario in which the photon or the $Z$ gauge boson are off-shell and found that one cannot expect an
increase of more than one order of magnitude of
the real part of the electromagnetic and weak
dipole moments. However,   an imaginary part can be developed provided that $s>2m_q$, which would
be about the smaller than the corresponding real part.

\acknowledgments{We acknowledge financial support from Conacyt and SNI (M\'exico). G.T.V would like
to thank partial support from VIEP-BUAP.}

\appendix

\section{Results in terms of Passarino-Veltman scalar functions}
\label{PassarinoResults}
As a cross-check for our calculation we have obtained results for the weak and
dipole moments by the
Passarino-Veltman reduction scheme. We will express our results in terms of two-point $B_0$   and
three-point $C_0$ scalar functions, which can be evaluated via the
numerical FF routines \cite{vanOldenborgh:1989wn,*Hahn:1998yk}.

\subsection{Fermion weak Dipole moments}

For the $F_a$ and $G_a$ functions appearing in the fermion WEDM and
WEDM of Eqs. (\ref{aWi}) and (\ref{dWi}) we obtain:
\begin{eqnarray}
F_1(z_1,z_2,z_3)&=&
\frac{2}{z_1 \left(4 z_1-z_3\right){}^2} \left(2   \left(z_1^2+\left(2 z_2-z_3+2\right) z_1-3
   \left(z_2-1\right){}^2+\left(z_2-2\right)
z_3\right)z_1C_1(z_1,z_2,z_3)\right.\nonumber\\&+&\left.\left(2
   z_1 \left(z_1-5 z_2+5\right)+\left(z_1+z_2-1\right) z_3\right)
   \Delta _1(z_1,z_2,z_3)+\left(4 z_1-z_3\right) \left(\left(z_2-1\right)
   \Delta _2(z_1,z_2,z_3)+z_1\right)\right),\\
F_2(z_1,z_2,z_3)&=&\frac{12 \left(z_1+z_2-1\right) }{\left(4
   z_1-z_3\right)^2} \left(\Delta
   _3(z_1,z_2,z_3)- \left(z_1+z_2-1\right)C_2(z_1,z_2,z_3)\right)-\frac{2}{z_1(z_3-4
z_1)}\left((1-z_2)\Delta
   _4(z_1,z_2,z_3)\right.\nonumber\\&-&\left.\left(z_1-z_2+1\right) \Delta
_3(z_1,z_2,z_3)+2 \left(z_1+2 z_2-1\right)
   z_1C_2(z_1,z_2,z_3)+z_1\right),\\
G_1(z_1,z_2,z_3)&=&
\frac{4}{4
   z_1-z_3} \left(\left(1+z_1-z_2\right)C_1(z_1,z_2,z_3) -\Delta _1(z_1,z_2,z_3)\right),\\
G_2(z_1,z_2,z_3)&=&
\frac{2}{4 z_1-z_3}\left(\left(2 (1+z_1-z_2)- z_3\right)C_2(z_1,z_2,z_3)+2 \Delta
_3(z_1,z_2,z_3)\right),
\end{eqnarray}
with the $\Delta_i$ and  $C_i$ functions  given in terms of scalar functions  (we  use the notation
of Ref. \cite{Mertig:1990an}) as
follows:
\begin{eqnarray}
\Delta_1(x,y,z)&=&B_0(x m_S^2, m_S^2, y m_S^2)- B_0(z m_S^2, y m_S^2, y m_S^2),\\
\Delta_2(x,y,z)&=& B_0(0, y m_S^2,  m_S^2)-B_0(z m_S^2, y m_S^2, y m_S^2),\\
\Delta_3(x,y,z)&=&B_0(x m_S^2, m_S^2, y m_S^2)-B_0(z m_S^2, m_S^2, m_S^2),\\
\Delta_4(x,y,z)&=&B_0(0, y m_S^2,  m_S^2)-B_0(z m_S^2, m_S^2, m_S^2),\\
C_1(x,y,z)&=&m_S^2 C_0(x m_S^2, x m_S^2, z m_S^2, m_S^2, y m_S^2, m_S^2),\\
C_2(x,y,z)&=&m_S^2 C_0(x m_S^2, x m_S^2, z m_S^2, y m_S^2, m_S^2, y m_S^2).
\end{eqnarray}
Notice that the $\Delta_i$ and $C_i$ functions are ultraviolet finite and independent of the
leptoquark mass. The  results for the $CP$-violating $d^W_\tau$ are in agreement with the results presented in Ref. \cite{Bernreuther:1996dr} when  $x_Z$ is replaced by $\frac{s}{m_Z^2}$.

\subsection{Fermion electromagnetic dipole moments}
For completeness we also present the results for the MDM and the EDM of a fermion in terms of scalar
functions. The
$F_a^\gamma$ and $G_a^\gamma$ functions of Eqs. (\ref{Fagfeyn}) and (\ref{Gagfeyn}) are given by

\begin{eqnarray}
F_1^\gamma(z_1,z_2)&=&-
\frac{1}{z_1^2 \lambda(z_1,z_2)}\left(2 z_2 \left(2 z_1+\lambda(z_1,z_2) \right)\Delta_5(z_1,z_2)-2
   \left(z_1 \left(1-z_1+z_2\right)+\lambda(z_1,z_2)
\right)\Delta_6(z_1,z_2)\right.\nonumber\\&+&\left.z_1 \left(4 z_2+\lambda(z_1,z_2) -4\right)+2
   \left(z_2-1\right) \lambda(z_1,z_2) +4 z_1^2\right),\\
F_2^\gamma(z_1,z_2)&=&-\frac{1}{z_1^2 \lambda(z_1,z_2) }
\left(2 z_2 \left(z_1
   \left(z_2+1\right)-\left(z_2-1\right){}^2\right)\Delta_5(z_1,z_2)+2
   \left(\left(z_2-1\right){}^2+\left(z_1-2\right)
   z_1\right)\Delta_6(z_1,z_2)\right.\nonumber\\&+&\left.z_1^3+\left(z_2-1\right)
\left(z_2+3\right) z_1-2
   \left(z_2-1\right){}^3\right),\\
G_1^\gamma(z_1,z_2)&=&-
\frac{1}{z_1 \lambda(z_1,z_2)}\left(2 \left(z_1^2-\left(2 z_2+1\right) z_1+\left(z_2-1\right)
   z_2\right)\Delta_5(z_1,z_2)+2 \left(z_1-z_2+1\right)\Delta_6(z_1,z_2)\right.\nonumber\\&+&\left.2
   \left(z_1-z_2+1\right){}^2\right),\\
G_2^\gamma(z_1,z_2)&=&-
\frac{1}{z_1 \lambda(z_1,z_2) }\left(2 \left(z_1-z_2+1\right) z_2\Delta_5(z_1,z_2)+2
\left(z_1+z_2-1\right)\Delta_6(z_1,z_2)+2
   \left(z_1^2-\left(z_2-1\right){}^2\right)\right),
\end{eqnarray}
with
\begin{eqnarray}
\label{PassVelFun2}
\Delta_5(x,y)&=&B_0(0, y\, m_S^2, y \,m_S^2)- B_0(x\, m_S^2, y \,m_S^2,
m_S^2),\\
\Delta_6(x,y)&=&B_0(0, m_S^2, m_S^2)- B_0(x\, m_S^2, y \,m_S^2, m_S^2),
\end{eqnarray}
and $\lambda(x,y)=\left(1+y-x\right)^2-4 y$.

\section{Lepton flavor violating decay $\ell_i \to \ell_k \gamma$}
\label{decaylitoljgamma}

We now present the results for the   decay amplitude (\ref{amplilkgamma}). There are only  contributions from the triangle diagrams of
Fig. \ref{litolkg}, whereas the bubble diagrams give rise to ultraviolet divergent terms that
violate electromagnetic gauge invariance and are exactly canceled out by  similar terms arising from
the triangle diagrams. The $I$ and $J$ functions appearing in the coefficients $A_L$ and $A_R$
of Eq. (\ref{ALlilkg}) are given by

\begin{eqnarray}
I(z_1,z_2,z_3)&=&Q_q I_1(z_1,z_2,z_3)+Q_SI_2(z_1,z_2,z_3),\\
J(z_1,z_2,z_3)&=&Q_q J_1(z_1,z_2,z_3)+Q_SJ_2(z_1,z_2,z_3),
\end{eqnarray}
with the $I_a$ and $J_a$ functions given by
\begin{eqnarray}
\label{Iafeyn}
I_a(z_1,z_2,z_3)=\frac{ \xi_a(x)}{(1-x)x
   \left(z_1-z_3\right){}^2} \left((x-1) x \left(z_1-z_3\right)-\left(x+(z_2-x z_3)(1-x)\right)
\log\left(\frac{\eta(x,z_1,z_2)}{\eta(x,z_3,z_2)}\right)\right),
\end{eqnarray}
and
\begin{eqnarray}
\label{Jafeyn}
J_a(z_1,z_2,z_3)=-\frac{ \xi_a(x) }{x \left(z_1-z_3\right)}
\log\left(\frac{\eta(x,z_1,z_2)}{\eta(x,z_3,z_2)}\right),
\end{eqnarray}
where $\xi_a(x)$ and $\eta(x,z_1,z_2)$ were defined after Eq. (\ref{Gaweak}). These equations also
reproduce the lepton MDM and EDM given in Eqs. (\ref{afi}) and (\ref{dfi}).

Finally, we present the
$I_a$ and $J_a$  functions in terms of Passarino-Veltman functions:
\begin{eqnarray}
   \label{I1ppas}
I_1(z_1,z_2,z_3)&=&\frac{1}{\left(z_1-z_3\right)^2}\left(\frac{1}{2 z_1}
   \left(\left(z_2-1\right) \left(z_3-2z_1\right)-z_1 z_3\right)\Delta_7(z_1,z_2)+\frac{1}{2}
   \left(z_2+z_3-1\right)\Delta_7(z_3,z_2)\right)\nonumber\\&+&\frac{1}{z_1-z_3}\left( \frac{1}{2
  } -z_2C_3(z_1,z_2,z_3)\right),\\
      \label{I2ppas}
I_2(z_1,z_2,z_3)&=&\frac{1}{\left(z_1-z_3\right)^2}\left(\frac{1}{2
   z_1 }
   \left(\left(1-z_2\right) \left(z_3-2z_1\right)-z_1z_3\right)\Delta_7(z_1,z_2)-\frac{1}{2}
    \left(z_2-z_3-1\right)\Delta_7(z_3,z_2)\right)\nonumber\\&+&\frac{1}{
   \left(z_3-z_1\right)}\left(\frac{1}{2}-C_4(z_1,z_2,z_3)\right),\\
\label{J1ppas}
J_1(z_1,z_2,z_3)&=&\frac{1}{z_1-z_3}\left(\Delta
   _5(z_3,z_2)+\Delta_7(z_1,z_2)\right)-C_3(z_1,z_2,z_3),\\
      \label{J2ppas}
J_2(z_1,z_2,z_3)&=& \frac{1}{z_1-z_3}\left(\Delta_7(z_1,z_2)-\Delta_7(z_3,z_2)\right),
   \end{eqnarray}
with
\begin{eqnarray}
\label{PassVelFun1}
\Delta_7(x,y)&=&B_0(x\, m_S^2, y \,m_S^2, m_S^2)- B_0(0, y\,m_S^2, m_S^2),\\
C_3(x,y,z)&=&m_S^2 C_0(x\, m_S^2, z\, m_S^2, 0, y\, m_S^2, m_S^2, y\, m_S^2),\\
C_4(x,y,z)&=&m_S^2 C_0(x\, m_S^2, z\, m_S^2, 0, m_S^2,
   y\, m_S^2, m_S^2).
\end{eqnarray}

\bibliographystyle{apsrev4-1}
\bibliography{Article}{}

\begin{thebibliography}{59}%
\makeatletter
\providecommand \@ifxundefined [1]{%
 \@ifx{#1\undefined}
}%
\providecommand \@ifnum [1]{%
 \ifnum #1\expandafter \@firstoftwo
 \else \expandafter \@secondoftwo
 \fi
}%
\providecommand \@ifx [1]{%
 \ifx #1\expandafter \@firstoftwo
 \else \expandafter \@secondoftwo
 \fi
}%
\providecommand \natexlab [1]{#1}%
\providecommand \enquote  [1]{``#1''}%
\providecommand \bibnamefont  [1]{#1}%
\providecommand \bibfnamefont [1]{#1}%
\providecommand \citenamefont [1]{#1}%
\providecommand \href@noop [0]{\@secondoftwo}%
\providecommand \href [0]{\begingroup \@sanitize@url \@href}%
\providecommand \@href[1]{\@@startlink{#1}\@@href}%
\providecommand \@@href[1]{\endgroup#1\@@endlink}%
\providecommand \@sanitize@url [0]{\catcode `\\12\catcode `\$12\catcode
  `\&12\catcode `\#12\catcode `\^12\catcode `\_12\catcode `\%12\relax}%
\providecommand \@@startlink[1]{}%
\providecommand \@@endlink[0]{}%
\providecommand \url  [0]{\begingroup\@sanitize@url \@url }%
\providecommand \@url [1]{\endgroup\@href {#1}{\urlprefix }}%
\providecommand \urlprefix  [0]{URL }%
\providecommand \Eprint [0]{\href }%
\providecommand \doibase [0]{http://dx.doi.org/}%
\providecommand \selectlanguage [0]{\@gobble}%
\providecommand \bibinfo  [0]{\@secondoftwo}%
\providecommand \bibfield  [0]{\@secondoftwo}%
\providecommand \translation [1]{[#1]}%
\providecommand \BibitemOpen [0]{}%
\providecommand \bibitemStop [0]{}%
\providecommand \bibitemNoStop [0]{.\EOS\space}%
\providecommand \EOS [0]{\spacefactor3000\relax}%
\providecommand \BibitemShut  [1]{\csname bibitem#1\endcsname}%
\let\auto@bib@innerbib\@empty
%</preamble>
\bibitem [{\citenamefont {Bennett}\ \emph {et~al.}(2006)\citenamefont {Bennett}
  \emph {et~al.}}]{Bennett:2006fi}%
  \BibitemOpen
  \bibfield  {author} {\bibinfo {author} {\bibfnamefont {G.}~\bibnamefont
  {Bennett}} \emph {et~al.} (\bibinfo {collaboration} {Muon G-2
  Collaboration}),\ }\href {\doibase 10.1103/PhysRevD.73.072003} {\bibfield
  {journal} {\bibinfo  {journal} {Phys.Rev.}\ }\textbf {\bibinfo {volume}
  {D73}},\ \bibinfo {pages} {072003} (\bibinfo {year} {2006})},\ \Eprint
  {http://arxiv.org/abs/hep-ex/0602035} {arXiv:hep-ex/0602035 [hep-ex]}
  \BibitemShut {NoStop}%
%%CITATION = HEP-EX/0602035;%%
\bibitem [{\citenamefont {Abdallah}\ \emph {et~al.}(2004)\citenamefont
  {Abdallah} \emph {et~al.}}]{Abdallah:2003xd}%
  \BibitemOpen
  \bibfield  {author} {\bibinfo {author} {\bibfnamefont {J.}~\bibnamefont
  {Abdallah}} \emph {et~al.} (\bibinfo {collaboration} {DELPHI
  Collaboration}),\ }\href@noop {} {\bibfield  {journal} {\bibinfo  {journal}
  {Eur.Phys.J.}\ }\textbf {\bibinfo {volume} {C35}},\ \bibinfo {pages} {159}
  (\bibinfo {year} {2004})},\ \Eprint {http://arxiv.org/abs/hep-ex/0406010}
  {arXiv:hep-ex/0406010 [hep-ex]} \BibitemShut {NoStop}%
%%CITATION = HEP-EX/0406010;%%
\bibitem [{\citenamefont {Eidelman}\ and\ \citenamefont
  {Passera}(2007)}]{Eidelman:2007sb}%
  \BibitemOpen
  \bibfield  {author} {\bibinfo {author} {\bibfnamefont {S.}~\bibnamefont
  {Eidelman}}\ and\ \bibinfo {author} {\bibfnamefont {M.}~\bibnamefont
  {Passera}},\ }\href {\doibase 10.1142/S0217732307022694} {\bibfield
  {journal} {\bibinfo  {journal} {Mod.Phys.Lett.}\ }\textbf {\bibinfo {volume}
  {A22}},\ \bibinfo {pages} {159} (\bibinfo {year} {2007})},\ \Eprint
  {http://arxiv.org/abs/hep-ph/0701260} {arXiv:hep-ph/0701260 [hep-ph]}
  \BibitemShut {NoStop}%
%%CITATION = HEP-PH/0701260;%%
\bibitem [{\citenamefont {Biggio}(2008)}]{Biggio:2008in}%
  \BibitemOpen
  \bibfield  {author} {\bibinfo {author} {\bibfnamefont {C.}~\bibnamefont
  {Biggio}},\ }\href {\doibase 10.1016/j.physletb.2008.09.004} {\bibfield
  {journal} {\bibinfo  {journal} {Phys.Lett.}\ }\textbf {\bibinfo {volume}
  {B668}},\ \bibinfo {pages} {378} (\bibinfo {year} {2008})},\ \Eprint
  {http://arxiv.org/abs/0806.2558} {arXiv:0806.2558 [hep-ph]} \BibitemShut
  {NoStop}%
%%CITATION = ARXIV:0806.2558;%%
\bibitem [{\citenamefont {Ibrahim}\ and\ \citenamefont
  {Nath}(2008)}]{Ibrahim:2008gg}%
  \BibitemOpen
  \bibfield  {author} {\bibinfo {author} {\bibfnamefont {T.}~\bibnamefont
  {Ibrahim}}\ and\ \bibinfo {author} {\bibfnamefont {P.}~\bibnamefont {Nath}},\
  }\href {\doibase 10.1103/PhysRevD.78.075013} {\bibfield  {journal} {\bibinfo
  {journal} {Phys.Rev.}\ }\textbf {\bibinfo {volume} {D78}},\ \bibinfo {pages}
  {075013} (\bibinfo {year} {2008})},\ \Eprint {http://arxiv.org/abs/0806.3880}
  {arXiv:0806.3880 [hep-ph]} \BibitemShut {NoStop}%
%%CITATION = ARXIV:0806.3880;%%
\bibitem [{\citenamefont {Moyotl}\ and\ \citenamefont
  {Tavares-Velasco}(2012)}]{Moyotl:2012zz}%
  \BibitemOpen
  \bibfield  {author} {\bibinfo {author} {\bibfnamefont {A.}~\bibnamefont
  {Moyotl}}\ and\ \bibinfo {author} {\bibfnamefont {G.}~\bibnamefont
  {Tavares-Velasco}},\ }\href {\doibase 10.1103/PhysRevD.86.013014} {\bibfield
  {journal} {\bibinfo  {journal} {Phys.Rev.}\ }\textbf {\bibinfo {volume}
  {D86}},\ \bibinfo {pages} {013014} (\bibinfo {year} {2012})},\ \Eprint
  {http://arxiv.org/abs/1210.1994} {arXiv:1210.1994 [hep-ph]} \BibitemShut
  {NoStop}%
%%CITATION = ARXIV:1210.1994;%%
\bibitem [{\citenamefont {Regan}\ \emph {et~al.}(2002)\citenamefont {Regan},
  \citenamefont {Commins}, \citenamefont {Schmidt},\ and\ \citenamefont
  {DeMille}}]{Regan:2002ta}%
  \BibitemOpen
  \bibfield  {author} {\bibinfo {author} {\bibfnamefont {B.}~\bibnamefont
  {Regan}}, \bibinfo {author} {\bibfnamefont {E.}~\bibnamefont {Commins}},
  \bibinfo {author} {\bibfnamefont {C.}~\bibnamefont {Schmidt}}, \ and\
  \bibinfo {author} {\bibfnamefont {D.}~\bibnamefont {DeMille}},\ }\href
  {\doibase 10.1103/PhysRevLett.88.071805} {\bibfield  {journal} {\bibinfo
  {journal} {Phys.Rev.Lett.}\ }\textbf {\bibinfo {volume} {88}},\ \bibinfo
  {pages} {071805} (\bibinfo {year} {2002})}\BibitemShut {NoStop}%
%%CITATION = PRLTA,88,071805;%%
\bibitem [{\citenamefont {Bennett}\ \emph {et~al.}(2009)\citenamefont {Bennett}
  \emph {et~al.}}]{Bennett:2008dy}%
  \BibitemOpen
  \bibfield  {author} {\bibinfo {author} {\bibfnamefont {G.}~\bibnamefont
  {Bennett}} \emph {et~al.} (\bibinfo {collaboration} {Muon (g-2)
  Collaboration}),\ }\href {\doibase 10.1103/PhysRevD.80.052008} {\bibfield
  {journal} {\bibinfo  {journal} {Phys.Rev.}\ }\textbf {\bibinfo {volume}
  {D80}},\ \bibinfo {pages} {052008} (\bibinfo {year} {2009})},\ \Eprint
  {http://arxiv.org/abs/0811.1207} {arXiv:0811.1207 [hep-ex]} \BibitemShut
  {NoStop}%
%%CITATION = ARXIV:0811.1207;%%
\bibitem [{\citenamefont {Inami}\ \emph {et~al.}(2003)\citenamefont {Inami}
  \emph {et~al.}}]{Inami:2002ah}%
  \BibitemOpen
  \bibfield  {author} {\bibinfo {author} {\bibfnamefont {K.}~\bibnamefont
  {Inami}} \emph {et~al.} (\bibinfo {collaboration} {Belle Collaboration}),\
  }\href {\doibase 10.1016/S0370-2693(02)02984-2} {\bibfield  {journal}
  {\bibinfo  {journal} {Phys.Lett.}\ }\textbf {\bibinfo {volume} {B551}},\
  \bibinfo {pages} {16} (\bibinfo {year} {2003})},\ \Eprint
  {http://arxiv.org/abs/hep-ex/0210066} {arXiv:hep-ex/0210066 [hep-ex]}
  \BibitemShut {NoStop}%
%%CITATION = HEP-EX/0210066;%%
\bibitem [{\citenamefont {Gonzalez-Sprinberg}\ \emph
  {et~al.}(2007)\citenamefont {Gonzalez-Sprinberg}, \citenamefont {Bernabeu},\
  and\ \citenamefont {Vidal}}]{GonzalezSprinberg:2007qj}%
  \BibitemOpen
  \bibfield  {author} {\bibinfo {author} {\bibfnamefont {G.}~\bibnamefont
  {Gonzalez-Sprinberg}}, \bibinfo {author} {\bibfnamefont {J.}~\bibnamefont
  {Bernabeu}}, \ and\ \bibinfo {author} {\bibfnamefont {J.}~\bibnamefont
  {Vidal}},\ }\href@noop {} {\  (\bibinfo {year} {2007})},\ \Eprint
  {http://arxiv.org/abs/0707.1658} {arXiv:0707.1658 [hep-ph]} \BibitemShut
  {NoStop}%
%%CITATION = ARXIV:0707.1658;%%
\bibitem [{\citenamefont {Bernabeu}\ \emph {et~al.}(2008)\citenamefont
  {Bernabeu}, \citenamefont {Gonzalez-Sprinberg}, \citenamefont
  {Papavassiliou},\ and\ \citenamefont {Vidal}}]{Bernabeu:2007rr}%
  \BibitemOpen
  \bibfield  {author} {\bibinfo {author} {\bibfnamefont {J.}~\bibnamefont
  {Bernabeu}}, \bibinfo {author} {\bibfnamefont {G.}~\bibnamefont
  {Gonzalez-Sprinberg}}, \bibinfo {author} {\bibfnamefont {J.}~\bibnamefont
  {Papavassiliou}}, \ and\ \bibinfo {author} {\bibfnamefont {J.}~\bibnamefont
  {Vidal}},\ }\href {\doibase 10.1016/j.nuclphysb.2007.09.001} {\bibfield
  {journal} {\bibinfo  {journal} {Nucl.Phys.}\ }\textbf {\bibinfo {volume}
  {B790}},\ \bibinfo {pages} {160} (\bibinfo {year} {2008})},\ \Eprint
  {http://arxiv.org/abs/0707.2496} {arXiv:0707.2496 [hep-ph]} \BibitemShut
  {NoStop}%
%%CITATION = ARXIV:0707.2496;%%
\bibitem [{\citenamefont {O'Leary}\ \emph {et~al.}(2010)\citenamefont {O'Leary}
  \emph {et~al.}}]{O'Leary:2010af}%
  \BibitemOpen
  \bibfield  {author} {\bibinfo {author} {\bibfnamefont {B.}~\bibnamefont
  {O'Leary}} \emph {et~al.} (\bibinfo {collaboration} {SuperB Collaboration}),\
  }\href@noop {} {\  (\bibinfo {year} {2010})},\ \Eprint
  {http://arxiv.org/abs/1008.1541} {arXiv:1008.1541 [hep-ex]} \BibitemShut
  {NoStop}%
%%CITATION = ARXIV:1008.1541;%%
\bibitem [{\citenamefont {Laursen}\ \emph {et~al.}(1984)\citenamefont
  {Laursen}, \citenamefont {Samuel},\ and\ \citenamefont
  {Sen}}]{Laursen:1983sm}%
  \BibitemOpen
  \bibfield  {author} {\bibinfo {author} {\bibfnamefont {M.}~\bibnamefont
  {Laursen}}, \bibinfo {author} {\bibfnamefont {M.~A.}\ \bibnamefont {Samuel}},
  \ and\ \bibinfo {author} {\bibfnamefont {A.}~\bibnamefont {Sen}},\ }\href
  {\doibase 10.1103/PhysRevD.29.2652, 10.1103/PhysRevD.56.3155} {\bibfield
  {journal} {\bibinfo  {journal} {Phys.Rev.}\ }\textbf {\bibinfo {volume}
  {D29}},\ \bibinfo {pages} {2652} (\bibinfo {year} {1984})}\BibitemShut
  {NoStop}%
%%CITATION = PHRVA,D29,2652;%%
\bibitem [{\citenamefont {Bernreuther}\ \emph {et~al.}(1989)\citenamefont
  {Bernreuther}, \citenamefont {Low}, \citenamefont {Ma},\ and\ \citenamefont
  {Nachtmann}}]{Bernreuther:1988jr}%
  \BibitemOpen
  \bibfield  {author} {\bibinfo {author} {\bibfnamefont {W.}~\bibnamefont
  {Bernreuther}}, \bibinfo {author} {\bibfnamefont {U.}~\bibnamefont {Low}},
  \bibinfo {author} {\bibfnamefont {J.}~\bibnamefont {Ma}}, \ and\ \bibinfo
  {author} {\bibfnamefont {O.}~\bibnamefont {Nachtmann}},\ }\href {\doibase
  10.1007/BF02430617} {\bibfield  {journal} {\bibinfo  {journal} {Z.Phys.}\
  }\textbf {\bibinfo {volume} {C43}},\ \bibinfo {pages} {117} (\bibinfo {year}
  {1989})}\BibitemShut {NoStop}%
%%CITATION = ZEPYA,C43,117;%%
\bibitem [{\citenamefont {Bernabeu}\ \emph {et~al.}(1994)\citenamefont
  {Bernabeu}, \citenamefont {Gonzalez-Sprinberg},\ and\ \citenamefont
  {Vidal}}]{Bernabeu:1993er}%
  \BibitemOpen
  \bibfield  {author} {\bibinfo {author} {\bibfnamefont {J.}~\bibnamefont
  {Bernabeu}}, \bibinfo {author} {\bibfnamefont {G.}~\bibnamefont
  {Gonzalez-Sprinberg}}, \ and\ \bibinfo {author} {\bibfnamefont
  {J.}~\bibnamefont {Vidal}},\ }\href {\doibase 10.1016/0370-2693(94)91209-2}
  {\bibfield  {journal} {\bibinfo  {journal} {Phys.Lett.}\ }\textbf {\bibinfo
  {volume} {B326}},\ \bibinfo {pages} {168} (\bibinfo {year}
  {1994})}\BibitemShut {NoStop}%
%%CITATION = PHLTA,B326,168;%%
\bibitem [{\citenamefont {Heister}\ \emph {et~al.}(2003)\citenamefont {Heister}
  \emph {et~al.}}]{Heister:2002ik}%
  \BibitemOpen
  \bibfield  {author} {\bibinfo {author} {\bibfnamefont {A.}~\bibnamefont
  {Heister}} \emph {et~al.} (\bibinfo {collaboration} {ALEPH Collaboration}),\
  }\href {\doibase 10.1140/epjc/s2003-01286-1} {\bibfield  {journal} {\bibinfo
  {journal} {Eur.Phys.J.}\ }\textbf {\bibinfo {volume} {C30}},\ \bibinfo
  {pages} {291} (\bibinfo {year} {2003})},\ \Eprint
  {http://arxiv.org/abs/hep-ex/0209066} {arXiv:hep-ex/0209066 [hep-ex]}
  \BibitemShut {NoStop}%
%%CITATION = HEP-EX/0209066;%%
\bibitem [{\citenamefont {Bernabeu}\ \emph {et~al.}(1995)\citenamefont
  {Bernabeu}, \citenamefont {Gonzalez-Sprinberg}, \citenamefont {Tung},\ and\
  \citenamefont {Vidal}}]{Bernabeu:1994wh}%
  \BibitemOpen
  \bibfield  {author} {\bibinfo {author} {\bibfnamefont {J.}~\bibnamefont
  {Bernabeu}}, \bibinfo {author} {\bibfnamefont {G.}~\bibnamefont
  {Gonzalez-Sprinberg}}, \bibinfo {author} {\bibfnamefont {M.}~\bibnamefont
  {Tung}}, \ and\ \bibinfo {author} {\bibfnamefont {J.}~\bibnamefont {Vidal}},\
  }\href {\doibase 10.1016/0550-3213(94)00525-J} {\bibfield  {journal}
  {\bibinfo  {journal} {Nucl.Phys.}\ }\textbf {\bibinfo {volume} {B436}},\
  \bibinfo {pages} {474} (\bibinfo {year} {1995})},\ \Eprint
  {http://arxiv.org/abs/hep-ph/9411289} {arXiv:hep-ph/9411289 [hep-ph]}
  \BibitemShut {NoStop}%
%%CITATION = HEP-PH/9411289;%%
\bibitem [{\citenamefont {Hayreter}\ and\ \citenamefont
  {Valencia}(2013)}]{Hayreter:2013vna}%
  \BibitemOpen
  \bibfield  {author} {\bibinfo {author} {\bibfnamefont {A.}~\bibnamefont
  {Hayreter}}\ and\ \bibinfo {author} {\bibfnamefont {G.}~\bibnamefont
  {Valencia}},\ }\href {\doibase 10.1103/PhysRevD.88.013015} {\bibfield
  {journal} {\bibinfo  {journal} {Phys.Rev.}\ }\textbf {\bibinfo {volume}
  {D88}},\ \bibinfo {pages} {013015} (\bibinfo {year} {2013})},\ \Eprint
  {http://arxiv.org/abs/1305.6833} {arXiv:1305.6833 [hep-ph]} \BibitemShut
  {NoStop}%
%%CITATION = ARXIV:1305.6833;%%
\bibitem [{\citenamefont {Queijeiro}(1994)}]{Queijeiro:1994kb}%
  \BibitemOpen
  \bibfield  {author} {\bibinfo {author} {\bibfnamefont {A.}~\bibnamefont
  {Queijeiro}},\ }\href {\doibase 10.1007/BF01557620} {\bibfield  {journal}
  {\bibinfo  {journal} {Z.Phys.}\ }\textbf {\bibinfo {volume} {C63}},\ \bibinfo
  {pages} {557} (\bibinfo {year} {1994})}\BibitemShut {NoStop}%
%%CITATION = ZEPYA,C63,557;%%
\bibitem [{\citenamefont {Bernabeu}\ \emph {et~al.}(1996)\citenamefont
  {Bernabeu}, \citenamefont {Comelli}, \citenamefont {Lavoura},\ and\
  \citenamefont {Silva}}]{Bernabeu:1995gs}%
  \BibitemOpen
  \bibfield  {author} {\bibinfo {author} {\bibfnamefont {J.}~\bibnamefont
  {Bernabeu}}, \bibinfo {author} {\bibfnamefont {D.}~\bibnamefont {Comelli}},
  \bibinfo {author} {\bibfnamefont {L.}~\bibnamefont {Lavoura}}, \ and\
  \bibinfo {author} {\bibfnamefont {J.~P.}\ \bibnamefont {Silva}},\ }\href
  {\doibase 10.1103/PhysRevD.53.5222} {\bibfield  {journal} {\bibinfo
  {journal} {Phys.Rev.}\ }\textbf {\bibinfo {volume} {D53}},\ \bibinfo {pages}
  {5222} (\bibinfo {year} {1996})},\ \Eprint
  {http://arxiv.org/abs/hep-ph/9509416} {arXiv:hep-ph/9509416 [hep-ph]}
  \BibitemShut {NoStop}%
%%CITATION = HEP-PH/9509416;%%
\bibitem [{\citenamefont {Gomez-Dumm}\ and\ \citenamefont
  {Gonzalez-Sprinberg}(1999)}]{GomezDumm:1999tz}%
  \BibitemOpen
  \bibfield  {author} {\bibinfo {author} {\bibfnamefont {D.}~\bibnamefont
  {Gomez-Dumm}}\ and\ \bibinfo {author} {\bibfnamefont {G.}~\bibnamefont
  {Gonzalez-Sprinberg}},\ }\href {\doibase 10.1007/s100520050633} {\bibfield
  {journal} {\bibinfo  {journal} {Eur.Phys.J.}\ }\textbf {\bibinfo {volume}
  {C11}},\ \bibinfo {pages} {293} (\bibinfo {year} {1999})},\ \Eprint
  {http://arxiv.org/abs/hep-ph/9905213} {arXiv:hep-ph/9905213 [hep-ph]}
  \BibitemShut {NoStop}%
%%CITATION = HEP-PH/9905213;%%
\bibitem [{\citenamefont {de~Carlos}\ and\ \citenamefont
  {Moreno}(1998)}]{deCarlos:1997br}%
  \BibitemOpen
  \bibfield  {author} {\bibinfo {author} {\bibfnamefont {B.}~\bibnamefont
  {de~Carlos}}\ and\ \bibinfo {author} {\bibfnamefont {J.}~\bibnamefont
  {Moreno}},\ }\href {\doibase 10.1016/S0550-3213(98)00033-9} {\bibfield
  {journal} {\bibinfo  {journal} {Nucl.Phys.}\ }\textbf {\bibinfo {volume}
  {B519}},\ \bibinfo {pages} {101} (\bibinfo {year} {1998})},\ \Eprint
  {http://arxiv.org/abs/hep-ph/9707487} {arXiv:hep-ph/9707487 [hep-ph]}
  \BibitemShut {NoStop}%
%%CITATION = HEP-PH/9707487;%%
\bibitem [{\citenamefont {Hollik}\ \emph
  {et~al.}(1998{\natexlab{a}})\citenamefont {Hollik}, \citenamefont {Illana},
  \citenamefont {Rigolin},\ and\ \citenamefont {Stockinger}}]{Hollik:1997vb}%
  \BibitemOpen
  \bibfield  {author} {\bibinfo {author} {\bibfnamefont {W.}~\bibnamefont
  {Hollik}}, \bibinfo {author} {\bibfnamefont {J.~I.}\ \bibnamefont {Illana}},
  \bibinfo {author} {\bibfnamefont {S.}~\bibnamefont {Rigolin}}, \ and\
  \bibinfo {author} {\bibfnamefont {D.}~\bibnamefont {Stockinger}},\ }\href
  {\doibase 10.1016/S0370-2693(97)01259-8} {\bibfield  {journal} {\bibinfo
  {journal} {Phys.Lett.}\ }\textbf {\bibinfo {volume} {B416}},\ \bibinfo
  {pages} {345} (\bibinfo {year} {1998}{\natexlab{a}})},\ \Eprint
  {http://arxiv.org/abs/hep-ph/9707437} {arXiv:hep-ph/9707437 [hep-ph]}
  \BibitemShut {NoStop}%
%%CITATION = HEP-PH/9707437;%%
\bibitem [{\citenamefont {Hollik}\ \emph
  {et~al.}(1998{\natexlab{b}})\citenamefont {Hollik}, \citenamefont {Illana},
  \citenamefont {Rigolin},\ and\ \citenamefont {Stockinger}}]{Hollik:1997ph}%
  \BibitemOpen
  \bibfield  {author} {\bibinfo {author} {\bibfnamefont {W.}~\bibnamefont
  {Hollik}}, \bibinfo {author} {\bibfnamefont {J.~I.}\ \bibnamefont {Illana}},
  \bibinfo {author} {\bibfnamefont {S.}~\bibnamefont {Rigolin}}, \ and\
  \bibinfo {author} {\bibfnamefont {D.}~\bibnamefont {Stockinger}},\ }\href
  {\doibase 10.1016/S0370-2693(98)00247-0} {\bibfield  {journal} {\bibinfo
  {journal} {Phys.Lett.}\ }\textbf {\bibinfo {volume} {B425}},\ \bibinfo
  {pages} {322} (\bibinfo {year} {1998}{\natexlab{b}})},\ \Eprint
  {http://arxiv.org/abs/hep-ph/9711322} {arXiv:hep-ph/9711322 [hep-ph]}
  \BibitemShut {NoStop}%
%%CITATION = HEP-PH/9711322;%%
\bibitem [{\citenamefont {Arnold}\ \emph {et~al.}(2013)\citenamefont {Arnold},
  \citenamefont {Fornal},\ and\ \citenamefont {Wise}}]{Arnold:2013cva}%
  \BibitemOpen
  \bibfield  {author} {\bibinfo {author} {\bibfnamefont {J.~M.}\ \bibnamefont
  {Arnold}}, \bibinfo {author} {\bibfnamefont {B.}~\bibnamefont {Fornal}}, \
  and\ \bibinfo {author} {\bibfnamefont {M.~B.}\ \bibnamefont {Wise}},\ }\href
  {\doibase 10.1103/PhysRevD.88.035009} {\bibfield  {journal} {\bibinfo
  {journal} {Phys.Rev.}\ }\textbf {\bibinfo {volume} {D88}},\ \bibinfo {pages}
  {035009} (\bibinfo {year} {2013})},\ \Eprint {http://arxiv.org/abs/1304.6119}
  {arXiv:1304.6119 [hep-ph]} \BibitemShut {NoStop}%
%%CITATION = ARXIV:1304.6119;%%
\bibitem [{\citenamefont {Dorsner}\ \emph {et~al.}(2009)\citenamefont
  {Dorsner}, \citenamefont {Fajfer}, \citenamefont {Kamenik},\ and\
  \citenamefont {Kosnik}}]{Dorsner:2009cu}%
  \BibitemOpen
  \bibfield  {author} {\bibinfo {author} {\bibfnamefont {I.}~\bibnamefont
  {Dorsner}}, \bibinfo {author} {\bibfnamefont {S.}~\bibnamefont {Fajfer}},
  \bibinfo {author} {\bibfnamefont {J.~F.}\ \bibnamefont {Kamenik}}, \ and\
  \bibinfo {author} {\bibfnamefont {N.}~\bibnamefont {Kosnik}},\ }\href
  {\doibase 10.1016/j.physletb.2009.10.087} {\bibfield  {journal} {\bibinfo
  {journal} {Phys.Lett.}\ }\textbf {\bibinfo {volume} {B682}},\ \bibinfo
  {pages} {67} (\bibinfo {year} {2009})},\ \Eprint
  {http://arxiv.org/abs/0906.5585} {arXiv:0906.5585 [hep-ph]} \BibitemShut
  {NoStop}%
%%CITATION = ARXIV:0906.5585;%%
\bibitem [{\citenamefont {Doršner}\ \emph {et~al.}(2013)\citenamefont
  {Doršner}, \citenamefont {Fajfer}, \citenamefont {Košnik},\ and\
  \citenamefont {Nišandžić}}]{Dorsner:2013tla}%
  \BibitemOpen
  \bibfield  {author} {\bibinfo {author} {\bibfnamefont {I.}~\bibnamefont
  {Doršner}}, \bibinfo {author} {\bibfnamefont {S.}~\bibnamefont {Fajfer}},
  \bibinfo {author} {\bibfnamefont {N.}~\bibnamefont {Košnik}}, \ and\
  \bibinfo {author} {\bibfnamefont {I.}~\bibnamefont {Nišandžić}},\ }\href
  {\doibase 10.1007/JHEP11(2013)084} {\bibfield  {journal} {\bibinfo  {journal}
  {JHEP}\ }\textbf {\bibinfo {volume} {1311}},\ \bibinfo {pages} {084}
  (\bibinfo {year} {2013})},\ \Eprint {http://arxiv.org/abs/1306.6493}
  {arXiv:1306.6493 [hep-ph]} \BibitemShut {NoStop}%
%%CITATION = ARXIV:1306.6493;%%
\bibitem [{\citenamefont {Pati}\ and\ \citenamefont
  {Salam}(1974)}]{Pati:1974yy}%
  \BibitemOpen
  \bibfield  {author} {\bibinfo {author} {\bibfnamefont {J.~C.}\ \bibnamefont
  {Pati}}\ and\ \bibinfo {author} {\bibfnamefont {A.}~\bibnamefont {Salam}},\
  }\href {\doibase 10.1103/PhysRevD.10.275, 10.1103/PhysRevD.11.703.2}
  {\bibfield  {journal} {\bibinfo  {journal} {Phys.Rev.}\ }\textbf {\bibinfo
  {volume} {D10}},\ \bibinfo {pages} {275} (\bibinfo {year}
  {1974})}\BibitemShut {NoStop}%
%%CITATION = PHRVA,D10,275;%%
\bibitem [{\citenamefont {Georgi}\ and\ \citenamefont
  {Glashow}(1974)}]{Georgi:1974sy}%
  \BibitemOpen
  \bibfield  {author} {\bibinfo {author} {\bibfnamefont {H.}~\bibnamefont
  {Georgi}}\ and\ \bibinfo {author} {\bibfnamefont {S.}~\bibnamefont
  {Glashow}},\ }\href {\doibase 10.1103/PhysRevLett.32.438} {\bibfield
  {journal} {\bibinfo  {journal} {Phys.Rev.Lett.}\ }\textbf {\bibinfo {volume}
  {32}},\ \bibinfo {pages} {438} (\bibinfo {year} {1974})}\BibitemShut
  {NoStop}%
%%CITATION = PRLTA,32,438;%%
\bibitem [{\citenamefont {Senjanovic}\ and\ \citenamefont
  {Sokorac}(1983)}]{Senjanovic:1982ex}%
  \BibitemOpen
  \bibfield  {author} {\bibinfo {author} {\bibfnamefont {G.}~\bibnamefont
  {Senjanovic}}\ and\ \bibinfo {author} {\bibfnamefont {A.}~\bibnamefont
  {Sokorac}},\ }\href {\doibase 10.1007/BF01574858} {\bibfield  {journal}
  {\bibinfo  {journal} {Z.Phys.}\ }\textbf {\bibinfo {volume} {C20}},\ \bibinfo
  {pages} {255} (\bibinfo {year} {1983})}\BibitemShut {NoStop}%
%%CITATION = ZEPYA,C20,255;%%
\bibitem [{\citenamefont {Frampton}\ and\ \citenamefont
  {Lee}(1990)}]{Frampton:1989fu}%
  \BibitemOpen
  \bibfield  {author} {\bibinfo {author} {\bibfnamefont {P.~H.}\ \bibnamefont
  {Frampton}}\ and\ \bibinfo {author} {\bibfnamefont {B.-H.}\ \bibnamefont
  {Lee}},\ }\href {\doibase 10.1103/PhysRevLett.64.619} {\bibfield  {journal}
  {\bibinfo  {journal} {Phys.Rev.Lett.}\ }\textbf {\bibinfo {volume} {64}},\
  \bibinfo {pages} {619} (\bibinfo {year} {1990})}\BibitemShut {NoStop}%
%%CITATION = PRLTA,64,619;%%
\bibitem [{\citenamefont {Schrempp}\ and\ \citenamefont
  {Schrempp}(1985)}]{Schrempp:1984nj}%
  \BibitemOpen
  \bibfield  {author} {\bibinfo {author} {\bibfnamefont {B.}~\bibnamefont
  {Schrempp}}\ and\ \bibinfo {author} {\bibfnamefont {F.}~\bibnamefont
  {Schrempp}},\ }\href {\doibase 10.1016/0370-2693(85)91450-9} {\bibfield
  {journal} {\bibinfo  {journal} {Phys.Lett.}\ }\textbf {\bibinfo {volume}
  {B153}},\ \bibinfo {pages} {101} (\bibinfo {year} {1985})}\BibitemShut
  {NoStop}%
%%CITATION = PHLTA,B153,101;%%
\bibitem [{\citenamefont {Farhi}\ and\ \citenamefont
  {Susskind}(1981)}]{Farhi:1980xs}%
  \BibitemOpen
  \bibfield  {author} {\bibinfo {author} {\bibfnamefont {E.}~\bibnamefont
  {Farhi}}\ and\ \bibinfo {author} {\bibfnamefont {L.}~\bibnamefont
  {Susskind}},\ }\href {\doibase 10.1016/0370-1573(81)90173-3} {\bibfield
  {journal} {\bibinfo  {journal} {Phys.Rept.}\ }\textbf {\bibinfo {volume}
  {74}},\ \bibinfo {pages} {277} (\bibinfo {year} {1981})}\BibitemShut
  {NoStop}%
%%CITATION = PRPLC,74,277;%%
\bibitem [{\citenamefont {Hill}\ and\ \citenamefont
  {Simmons}(2003)}]{Hill:2002ap}%
  \BibitemOpen
  \bibfield  {author} {\bibinfo {author} {\bibfnamefont {C.~T.}\ \bibnamefont
  {Hill}}\ and\ \bibinfo {author} {\bibfnamefont {E.~H.}\ \bibnamefont
  {Simmons}},\ }\href {\doibase 10.1016/S0370-1573(03)00140-6} {\bibfield
  {journal} {\bibinfo  {journal} {Phys.Rept.}\ }\textbf {\bibinfo {volume}
  {381}},\ \bibinfo {pages} {235} (\bibinfo {year} {2003})},\ \Eprint
  {http://arxiv.org/abs/hep-ph/0203079} {arXiv:hep-ph/0203079 [hep-ph]}
  \BibitemShut {NoStop}%
%%CITATION = HEP-PH/0203079;%%
\bibitem [{\citenamefont {Witten}(1985)}]{Witten:1985xc}%
  \BibitemOpen
  \bibfield  {author} {\bibinfo {author} {\bibfnamefont {E.}~\bibnamefont
  {Witten}},\ }\href {\doibase 10.1016/0550-3213(85)90603-0} {\bibfield
  {journal} {\bibinfo  {journal} {Nucl.Phys.}\ }\textbf {\bibinfo {volume}
  {B258}},\ \bibinfo {pages} {75} (\bibinfo {year} {1985})}\BibitemShut
  {NoStop}%
%%CITATION = NUPHA,B258,75;%%
\bibitem [{\citenamefont {Hewett}\ and\ \citenamefont
  {Rizzo}(1989)}]{Hewett:1988xc}%
  \BibitemOpen
  \bibfield  {author} {\bibinfo {author} {\bibfnamefont {J.~L.}\ \bibnamefont
  {Hewett}}\ and\ \bibinfo {author} {\bibfnamefont {T.~G.}\ \bibnamefont
  {Rizzo}},\ }\href {\doibase 10.1016/0370-1573(89)90071-9} {\bibfield
  {journal} {\bibinfo  {journal} {Phys.Rept.}\ }\textbf {\bibinfo {volume}
  {183}},\ \bibinfo {pages} {193} (\bibinfo {year} {1989})}\BibitemShut
  {NoStop}%
%%CITATION = PRPLC,183,193;%%
\bibitem [{\citenamefont {Davidson}\ \emph {et~al.}(1994)\citenamefont
  {Davidson}, \citenamefont {Bailey},\ and\ \citenamefont
  {Campbell}}]{Davidson:1993qk}%
  \BibitemOpen
  \bibfield  {author} {\bibinfo {author} {\bibfnamefont {S.}~\bibnamefont
  {Davidson}}, \bibinfo {author} {\bibfnamefont {D.~C.}\ \bibnamefont
  {Bailey}}, \ and\ \bibinfo {author} {\bibfnamefont {B.~A.}\ \bibnamefont
  {Campbell}},\ }\href {\doibase 10.1007/BF01552629} {\bibfield  {journal}
  {\bibinfo  {journal} {Z.Phys.}\ }\textbf {\bibinfo {volume} {C61}},\ \bibinfo
  {pages} {613} (\bibinfo {year} {1994})},\ \Eprint
  {http://arxiv.org/abs/hep-ph/9309310} {arXiv:hep-ph/9309310 [hep-ph]}
  \BibitemShut {NoStop}%
%%CITATION = HEP-PH/9309310;%%
\bibitem [{\citenamefont {Hewett}\ and\ \citenamefont
  {Rizzo}(1997)}]{Hewett:1997ce}%
  \BibitemOpen
  \bibfield  {author} {\bibinfo {author} {\bibfnamefont {J.~L.}\ \bibnamefont
  {Hewett}}\ and\ \bibinfo {author} {\bibfnamefont {T.~G.}\ \bibnamefont
  {Rizzo}},\ }\href {\doibase 10.1103/PhysRevD.56.5709} {\bibfield  {journal}
  {\bibinfo  {journal} {Phys.Rev.}\ }\textbf {\bibinfo {volume} {D56}},\
  \bibinfo {pages} {5709} (\bibinfo {year} {1997})},\ \Eprint
  {http://arxiv.org/abs/hep-ph/9703337} {arXiv:hep-ph/9703337 [hep-ph]}
  \BibitemShut {NoStop}%
%%CITATION = HEP-PH/9703337;%%
\bibitem [{\citenamefont {Buchmuller}\ \emph {et~al.}(1987)\citenamefont
  {Buchmuller}, \citenamefont {Ruckl},\ and\ \citenamefont
  {Wyler}}]{Buchmuller:1986zs}%
  \BibitemOpen
  \bibfield  {author} {\bibinfo {author} {\bibfnamefont {W.}~\bibnamefont
  {Buchmuller}}, \bibinfo {author} {\bibfnamefont {R.}~\bibnamefont {Ruckl}}, \
  and\ \bibinfo {author} {\bibfnamefont {D.}~\bibnamefont {Wyler}},\ }\href
  {\doibase 10.1016/0370-2693(87)90637-X} {\bibfield  {journal} {\bibinfo
  {journal} {Phys.Lett.}\ }\textbf {\bibinfo {volume} {B191}},\ \bibinfo
  {pages} {442} (\bibinfo {year} {1987})}\BibitemShut {NoStop}%
%%CITATION = PHLTA,B191,442;%%
\bibitem [{\citenamefont {Bernreuther}\ \emph {et~al.}(1997)\citenamefont
  {Bernreuther}, \citenamefont {Brandenburg},\ and\ \citenamefont
  {Overmann}}]{Bernreuther:1996dr}%
  \BibitemOpen
  \bibfield  {author} {\bibinfo {author} {\bibfnamefont {W.}~\bibnamefont
  {Bernreuther}}, \bibinfo {author} {\bibfnamefont {A.}~\bibnamefont
  {Brandenburg}}, \ and\ \bibinfo {author} {\bibfnamefont {P.}~\bibnamefont
  {Overmann}},\ }\href {\doibase 10.1016/S0370-2693(96)01501-8} {\bibfield
  {journal} {\bibinfo  {journal} {Phys.Lett.}\ }\textbf {\bibinfo {volume}
  {B391}},\ \bibinfo {pages} {413} (\bibinfo {year} {1997})},\ \Eprint
  {http://arxiv.org/abs/hep-ph/9608364} {arXiv:hep-ph/9608364 [hep-ph]}
  \BibitemShut {NoStop}%
%%CITATION = HEP-PH/9608364;%%
\bibitem [{\citenamefont {Poulose}\ and\ \citenamefont
  {Rindani}(1998)}]{Poulose:1997kt}%
  \BibitemOpen
  \bibfield  {author} {\bibinfo {author} {\bibfnamefont {P.}~\bibnamefont
  {Poulose}}\ and\ \bibinfo {author} {\bibfnamefont {S.~D.}\ \bibnamefont
  {Rindani}},\ }\href {\doibase 10.1007/BF02828932} {\bibfield  {journal}
  {\bibinfo  {journal} {Pramana}\ }\textbf {\bibinfo {volume} {51}},\ \bibinfo
  {pages} {387} (\bibinfo {year} {1998})},\ \Eprint
  {http://arxiv.org/abs/hep-ph/9708332} {arXiv:hep-ph/9708332 [hep-ph]}
  \BibitemShut {NoStop}%
%%CITATION = HEP-PH/9708332;%%
\bibitem [{\citenamefont {Mizukoshi}\ \emph {et~al.}(1995)\citenamefont
  {Mizukoshi}, \citenamefont {Eboli},\ and\ \citenamefont
  {Gonzalez-Garcia}}]{Mizukoshi:1994zy}%
  \BibitemOpen
  \bibfield  {author} {\bibinfo {author} {\bibfnamefont {J.}~\bibnamefont
  {Mizukoshi}}, \bibinfo {author} {\bibfnamefont {O.~J.}\ \bibnamefont
  {Eboli}}, \ and\ \bibinfo {author} {\bibfnamefont {M.}~\bibnamefont
  {Gonzalez-Garcia}},\ }\href {\doibase 10.1016/0550-3213(95)00162-L}
  {\bibfield  {journal} {\bibinfo  {journal} {Nucl.Phys.}\ }\textbf {\bibinfo
  {volume} {B443}},\ \bibinfo {pages} {20} (\bibinfo {year} {1995})},\ \Eprint
  {http://arxiv.org/abs/hep-ph/9411392} {arXiv:hep-ph/9411392 [hep-ph]}
  \BibitemShut {NoStop}%
%%CITATION = HEP-PH/9411392;%%
\bibitem [{\citenamefont {Djouadi}\ \emph {et~al.}(1990)\citenamefont
  {Djouadi}, \citenamefont {Kohler}, \citenamefont {Spira},\ and\ \citenamefont
  {Tutas}}]{Djouadi:1989md}%
  \BibitemOpen
  \bibfield  {author} {\bibinfo {author} {\bibfnamefont {A.}~\bibnamefont
  {Djouadi}}, \bibinfo {author} {\bibfnamefont {T.}~\bibnamefont {Kohler}},
  \bibinfo {author} {\bibfnamefont {M.}~\bibnamefont {Spira}}, \ and\ \bibinfo
  {author} {\bibfnamefont {J.}~\bibnamefont {Tutas}},\ }\href {\doibase
  10.1007/BF01560270} {\bibfield  {journal} {\bibinfo  {journal} {Z.Phys.}\
  }\textbf {\bibinfo {volume} {C46}},\ \bibinfo {pages} {679} (\bibinfo {year}
  {1990})}\BibitemShut {NoStop}%
%%CITATION = ZEPYA,C46,679;%%
\bibitem [{\citenamefont {Cheung}(2001)}]{Cheung:2001ip}%
  \BibitemOpen
  \bibfield  {author} {\bibinfo {author} {\bibfnamefont {K.-m.}\ \bibnamefont
  {Cheung}},\ }\href {\doibase 10.1103/PhysRevD.64.033001} {\bibfield
  {journal} {\bibinfo  {journal} {Phys.Rev.}\ }\textbf {\bibinfo {volume}
  {D64}},\ \bibinfo {pages} {033001} (\bibinfo {year} {2001})},\ \Eprint
  {http://arxiv.org/abs/hep-ph/0102238} {arXiv:hep-ph/0102238 [hep-ph]}
  \BibitemShut {NoStop}%
%%CITATION = HEP-PH/0102238;%%
\bibitem [{\citenamefont {Bernreuther}\ and\ \citenamefont
  {Suzuki}(1991)}]{Bernreuther:1990jx}%
  \BibitemOpen
  \bibfield  {author} {\bibinfo {author} {\bibfnamefont {W.}~\bibnamefont
  {Bernreuther}}\ and\ \bibinfo {author} {\bibfnamefont {M.}~\bibnamefont
  {Suzuki}},\ }\href {\doibase 10.1103/RevModPhys.63.313} {\bibfield  {journal}
  {\bibinfo  {journal} {Rev.Mod.Phys.}\ }\textbf {\bibinfo {volume} {63}},\
  \bibinfo {pages} {313} (\bibinfo {year} {1991})}\BibitemShut {NoStop}%
%%CITATION = RMPHA,63,313;%%
\bibitem [{\citenamefont {Shanker}(1982)}]{Shanker:1982nd}%
  \BibitemOpen
  \bibfield  {author} {\bibinfo {author} {\bibfnamefont {O.~U.}\ \bibnamefont
  {Shanker}},\ }\href {\doibase 10.1016/0550-3213(82)90196-1} {\bibfield
  {journal} {\bibinfo  {journal} {Nucl.Phys.}\ }\textbf {\bibinfo {volume}
  {B204}},\ \bibinfo {pages} {375} (\bibinfo {year} {1982})}\BibitemShut
  {NoStop}%
%%CITATION = NUPHA,B204,375;%%
\bibitem [{\citenamefont {Chatrchyan}\ \emph {et~al.}(2012)\citenamefont
  {Chatrchyan} \emph {et~al.}}]{Chatrchyan:2012st}%
  \BibitemOpen
  \bibfield  {author} {\bibinfo {author} {\bibfnamefont {S.}~\bibnamefont
  {Chatrchyan}} \emph {et~al.} (\bibinfo {collaboration} {CMS Collaboration}),\
  }\href {\doibase 10.1007/JHEP12(2012)055} {\bibfield  {journal} {\bibinfo
  {journal} {JHEP}\ }\textbf {\bibinfo {volume} {1212}},\ \bibinfo {pages}
  {055} (\bibinfo {year} {2012})},\ \Eprint {http://arxiv.org/abs/1210.5627}
  {arXiv:1210.5627 [hep-ex]} \BibitemShut {NoStop}%
%%CITATION = ARXIV:1210.5627;%%
\bibitem [{\citenamefont {Keith}\ and\ \citenamefont
  {Ma}(1997)}]{Keith:1997fv}%
  \BibitemOpen
  \bibfield  {author} {\bibinfo {author} {\bibfnamefont {E.}~\bibnamefont
  {Keith}}\ and\ \bibinfo {author} {\bibfnamefont {E.}~\bibnamefont {Ma}},\
  }\href {\doibase 10.1103/PhysRevLett.79.4318} {\bibfield  {journal} {\bibinfo
   {journal} {Phys.Rev.Lett.}\ }\textbf {\bibinfo {volume} {79}},\ \bibinfo
  {pages} {4318} (\bibinfo {year} {1997})},\ \Eprint
  {http://arxiv.org/abs/hep-ph/9707214} {arXiv:hep-ph/9707214 [hep-ph]}
  \BibitemShut {NoStop}%
%%CITATION = HEP-PH/9707214;%%
\bibitem [{\citenamefont {Benbrik}\ and\ \citenamefont
  {Chua}(2008)}]{Benbrik:2008si}%
  \BibitemOpen
  \bibfield  {author} {\bibinfo {author} {\bibfnamefont {R.}~\bibnamefont
  {Benbrik}}\ and\ \bibinfo {author} {\bibfnamefont {C.-K.}\ \bibnamefont
  {Chua}},\ }\href {\doibase 10.1103/PhysRevD.78.075025} {\bibfield  {journal}
  {\bibinfo  {journal} {Phys.Rev.}\ }\textbf {\bibinfo {volume} {D78}},\
  \bibinfo {pages} {075025} (\bibinfo {year} {2008})},\ \Eprint
  {http://arxiv.org/abs/0807.4240} {arXiv:0807.4240 [hep-ph]} \BibitemShut
  {NoStop}%
%%CITATION = ARXIV:0807.4240;%%
\bibitem [{\citenamefont {Povarov}\ and\ \citenamefont
  {Smirnov}(2011)}]{Povarov:2011zz}%
  \BibitemOpen
  \bibfield  {author} {\bibinfo {author} {\bibfnamefont {A.}~\bibnamefont
  {Povarov}}\ and\ \bibinfo {author} {\bibfnamefont {A.}~\bibnamefont
  {Smirnov}},\ }\href {\doibase 10.1134/S1063778811050188} {\bibfield
  {journal} {\bibinfo  {journal} {Phys.Atom.Nucl.}\ }\textbf {\bibinfo {volume}
  {74}},\ \bibinfo {pages} {732} (\bibinfo {year} {2011})}\BibitemShut
  {NoStop}%
%%CITATION = PANUE,74,732;%%
\bibitem [{\citenamefont {Adam}\ \emph {et~al.}(2013)\citenamefont {Adam} \emph
  {et~al.}}]{Adam:2013mnn}%
  \BibitemOpen
  \bibfield  {author} {\bibinfo {author} {\bibfnamefont {J.}~\bibnamefont
  {Adam}} \emph {et~al.} (\bibinfo {collaboration} {MEG Collaboration}),\
  }\href@noop {} {\  (\bibinfo {year} {2013})},\ \Eprint
  {http://arxiv.org/abs/1303.0754} {arXiv:1303.0754 [hep-ex]} \BibitemShut
  {NoStop}%
%%CITATION = ARXIV:1303.0754;%%
\bibitem [{\citenamefont {Aubert}\ \emph {et~al.}(2010)\citenamefont {Aubert}
  \emph {et~al.}}]{Aubert:2009ag}%
  \BibitemOpen
  \bibfield  {author} {\bibinfo {author} {\bibfnamefont {B.}~\bibnamefont
  {Aubert}} \emph {et~al.} (\bibinfo {collaboration} {BaBar Collaboration}),\
  }\href {\doibase 10.1103/PhysRevLett.104.021802} {\bibfield  {journal}
  {\bibinfo  {journal} {Phys.Rev.Lett.}\ }\textbf {\bibinfo {volume} {104}},\
  \bibinfo {pages} {021802} (\bibinfo {year} {2010})},\ \Eprint
  {http://arxiv.org/abs/0908.2381} {arXiv:0908.2381 [hep-ex]} \BibitemShut
  {NoStop}%
%%CITATION = ARXIV:0908.2381;%%
\bibitem [{\citenamefont {Bhattacharyya}\ \emph {et~al.}(1994)\citenamefont
  {Bhattacharyya}, \citenamefont {Ellis},\ and\ \citenamefont
  {Sridhar}}]{Bhattacharyya:1994ig}%
  \BibitemOpen
  \bibfield  {author} {\bibinfo {author} {\bibfnamefont {G.}~\bibnamefont
  {Bhattacharyya}}, \bibinfo {author} {\bibfnamefont {J.~R.}\ \bibnamefont
  {Ellis}}, \ and\ \bibinfo {author} {\bibfnamefont {K.}~\bibnamefont
  {Sridhar}},\ }\href {\doibase 10.1016/0370-2693(94)00927-9} {\bibfield
  {journal} {\bibinfo  {journal} {Phys.Lett.}\ }\textbf {\bibinfo {volume}
  {B336}},\ \bibinfo {pages} {100} (\bibinfo {year} {1994})},\ \Eprint
  {http://arxiv.org/abs/hep-ph/9406354} {arXiv:hep-ph/9406354 [hep-ph]}
  \BibitemShut {NoStop}%
%%CITATION = HEP-PH/9406354;%%
\bibitem [{\citenamefont {Papavassiliou}\ and\ \citenamefont
  {Parrinello}(1994)}]{Papavassiliou:1993qe}%
  \BibitemOpen
  \bibfield  {author} {\bibinfo {author} {\bibfnamefont {J.}~\bibnamefont
  {Papavassiliou}}\ and\ \bibinfo {author} {\bibfnamefont {C.}~\bibnamefont
  {Parrinello}},\ }\href {\doibase 10.1103/PhysRevD.50.3059} {\bibfield
  {journal} {\bibinfo  {journal} {Phys.Rev.}\ }\textbf {\bibinfo {volume}
  {D50}},\ \bibinfo {pages} {3059} (\bibinfo {year} {1994})},\ \Eprint
  {http://arxiv.org/abs/hep-ph/9311284} {arXiv:hep-ph/9311284 [hep-ph]}
  \BibitemShut {NoStop}%
%%CITATION = HEP-PH/9311284;%%
\bibitem [{\citenamefont {Denner}\ \emph {et~al.}(1994)\citenamefont {Denner},
  \citenamefont {Weiglein},\ and\ \citenamefont {Dittmaier}}]{Denner:1994nn}%
  \BibitemOpen
  \bibfield  {author} {\bibinfo {author} {\bibfnamefont {A.}~\bibnamefont
  {Denner}}, \bibinfo {author} {\bibfnamefont {G.}~\bibnamefont {Weiglein}}, \
  and\ \bibinfo {author} {\bibfnamefont {S.}~\bibnamefont {Dittmaier}},\ }\href
  {\doibase 10.1016/0370-2693(94)90162-7} {\bibfield  {journal} {\bibinfo
  {journal} {Phys.Lett.}\ }\textbf {\bibinfo {volume} {B333}},\ \bibinfo
  {pages} {420} (\bibinfo {year} {1994})},\ \Eprint
  {http://arxiv.org/abs/hep-ph/9406204} {arXiv:hep-ph/9406204 [hep-ph]}
  \BibitemShut {NoStop}%
%%CITATION = HEP-PH/9406204;%%
\bibitem [{\citenamefont {Ibrahim}\ and\ \citenamefont
  {Nath}(2010)}]{Ibrahim:2010va}%
  \BibitemOpen
  \bibfield  {author} {\bibinfo {author} {\bibfnamefont {T.}~\bibnamefont
  {Ibrahim}}\ and\ \bibinfo {author} {\bibfnamefont {P.}~\bibnamefont {Nath}},\
  }\href {\doibase 10.1103/PhysRevD.81.033007} {\bibfield  {journal} {\bibinfo
  {journal} {Phys.Rev.}\ }\textbf {\bibinfo {volume} {D81}},\ \bibinfo {pages}
  {033007} (\bibinfo {year} {2010})},\ \Eprint {http://arxiv.org/abs/1001.0231}
  {arXiv:1001.0231 [hep-ph]} \BibitemShut {NoStop}%
%%CITATION = ARXIV:1001.0231;%%
\bibitem [{\citenamefont {van Oldenborgh}\ and\ \citenamefont
  {Vermaseren}(1990)}]{vanOldenborgh:1989wn}%
  \BibitemOpen
  \bibfield  {author} {\bibinfo {author} {\bibfnamefont {G.}~\bibnamefont {van
  Oldenborgh}}\ and\ \bibinfo {author} {\bibfnamefont {J.}~\bibnamefont
  {Vermaseren}},\ }\href {\doibase 10.1007/BF01621031} {\bibfield  {journal}
  {\bibinfo  {journal} {Z.Phys.}\ }\textbf {\bibinfo {volume} {C46}},\ \bibinfo
  {pages} {425} (\bibinfo {year} {1990})}\BibitemShut {NoStop}%
%%CITATION = ZEPYA,C46,425;%%
\bibitem [{\citenamefont {Hahn}\ and\ \citenamefont
  {Perez-Victoria}(1999)}]{Hahn:1998yk}%
  \BibitemOpen
  \bibfield  {author} {\bibinfo {author} {\bibfnamefont {T.}~\bibnamefont
  {Hahn}}\ and\ \bibinfo {author} {\bibfnamefont {M.}~\bibnamefont
  {Perez-Victoria}},\ }\href {\doibase 10.1016/S0010-4655(98)00173-8}
  {\bibfield  {journal} {\bibinfo  {journal} {Comput.Phys.Commun.}\ }\textbf
  {\bibinfo {volume} {118}},\ \bibinfo {pages} {153} (\bibinfo {year}
  {1999})},\ \Eprint {http://arxiv.org/abs/hep-ph/9807565}
  {arXiv:hep-ph/9807565 [hep-ph]} \BibitemShut {NoStop}%
%%CITATION = HEP-PH/9807565;%%
\bibitem [{\citenamefont {Mertig}\ \emph {et~al.}(1991)\citenamefont {Mertig},
  \citenamefont {Bohm},\ and\ \citenamefont {Denner}}]{Mertig:1990an}%
  \BibitemOpen
  \bibfield  {author} {\bibinfo {author} {\bibfnamefont {R.}~\bibnamefont
  {Mertig}}, \bibinfo {author} {\bibfnamefont {M.}~\bibnamefont {Bohm}}, \ and\
  \bibinfo {author} {\bibfnamefont {A.}~\bibnamefont {Denner}},\ }\href
  {\doibase 10.1016/0010-4655(91)90130-D} {\bibfield  {journal} {\bibinfo
  {journal} {Comput.Phys.Commun.}\ }\textbf {\bibinfo {volume} {64}},\ \bibinfo
  {pages} {345} (\bibinfo {year} {1991})}\BibitemShut {NoStop}%
%%CITATION = CPHCB,64,345;%%
\end{thebibliography}%

\end{document}